\documentclass[%
 aip,
 rsi,%
 amsmath,amssymb,
]{revtex4-1}

\usepackage{graphicx}% Include figure files
\usepackage{dcolumn}% Align table columns on decimal point
\usepackage{bm}% bold math
\begin{document}

%\preprint{AIP/123-QED}
\title{A Cleanroom in a Glovebox}% Force line breaks with \\
%\thanks{Footnote to title of article.}

\author{Mason J. Gray}
\author{Narendra Kumar}
\author{Ryan O'Connor}%
\author{Marcel Hoek}
\author{Erin Sheridan}
\author{Meaghan C. Doyle}
\author{Marisa L. Romanelli}
\author{Gavin B. Osterhoudt}
\author{Yiping Wang}
\author{Vincent Plisson}
\affiliation{Department of Physics, Boston College, Chestnut Hill, MA. 02467, USA}
\author{Shiming Lei}
\affiliation{Department of Chemistry, Princeton University, Princeton, NJ. 08544, USA}
\author{Ruidan Zhong}
\affiliation{Condensed Matter Physics and Materials Science Department, Brookhaven National Laboratory, Upton, NY. 11973, USA}
\author{Bryan Rachmilowitz}
\author{He Zhao}
\affiliation{Department of Physics, Boston College, Chestnut Hill, MA. 02467, USA}
\author{Hikari Kitadai}
\affiliation{Department of Chemistry, Boston University, Boston, MA. 02215, USA}
\author{Steven Shepard}
\affiliation{Integrated Sciences Cleanroom Facility, Boston College, Chestnut Hill, MA. 02467, USA}
\author{Leslie M. Schoop}
\affiliation{Department of Chemistry, Princeton University, Princeton, NJ. 08544, USA}
\author{G. D. Gu}
\affiliation{Condensed Matter Physics and Materials Science Department, Brookhaven National Laboratory, Upton, NY. 11973, USA}
\author{Ilija Zeljkovic}
\affiliation{Department of Physics, Boston College, Chestnut Hill, MA. 02467, USA}
\author{Xi Ling}
\affiliation{Department of Chemistry, Boston University, Boston, MA. 02215, USA}
\affiliation{Division of Materials Science and Engineering, Boston University, Boston, MA, 02215, USA}
\affiliation{The Photonics Center, Boston University, Boston, MA. 02215, USA}
% Add more Authors here as I insert their stuff.

\author{K.S. Burch}
 \homepage{Corresponding author}
 \email{ks.burch@bc.edu}
\affiliation{Department of Physics, Boston College, Chestnut Hill, MA. 02467, USA}%\\This line break forced% with \\

\date{\today}% It is always \today, today,

%Use showkeys class option if keyword
\maketitle

The exploration of new materials, novel quantum phases, and devices requires ways to prepare cleaner samples with smaller feature sizes. Initially, this meant the use of a cleanroom that limits the amount and size of dust particles. However, many materials are highly sensitive to oxygen and water in the air. Furthermore, the ever-increasing demand for a quantum workforce, trained and able to use the equipment for creating and characterizing materials, calls for a dramatic reduction in the cost to create and operate such facilities. To this end, we present our cleanroom-in-a-glovebox, a system which allows for the fabrication and characterization of devices in an inert argon atmosphere. We demonstrate the ability to perform a wide range of characterization as well as fabrication steps, without the need for a dedicated room, all in an argon environment. Connection to a vacuum suitcase is also demonstrated to enable receiving from and transfer to various ultra-high vacuum (UHV) equipment including molecular-beam epitaxy (MBE) and scanning tunneling microscopy (STM). 
\newpage

\section{\label{sec:level1}Introduction}
\begin{figure}[h]
    \centering
    \includegraphics{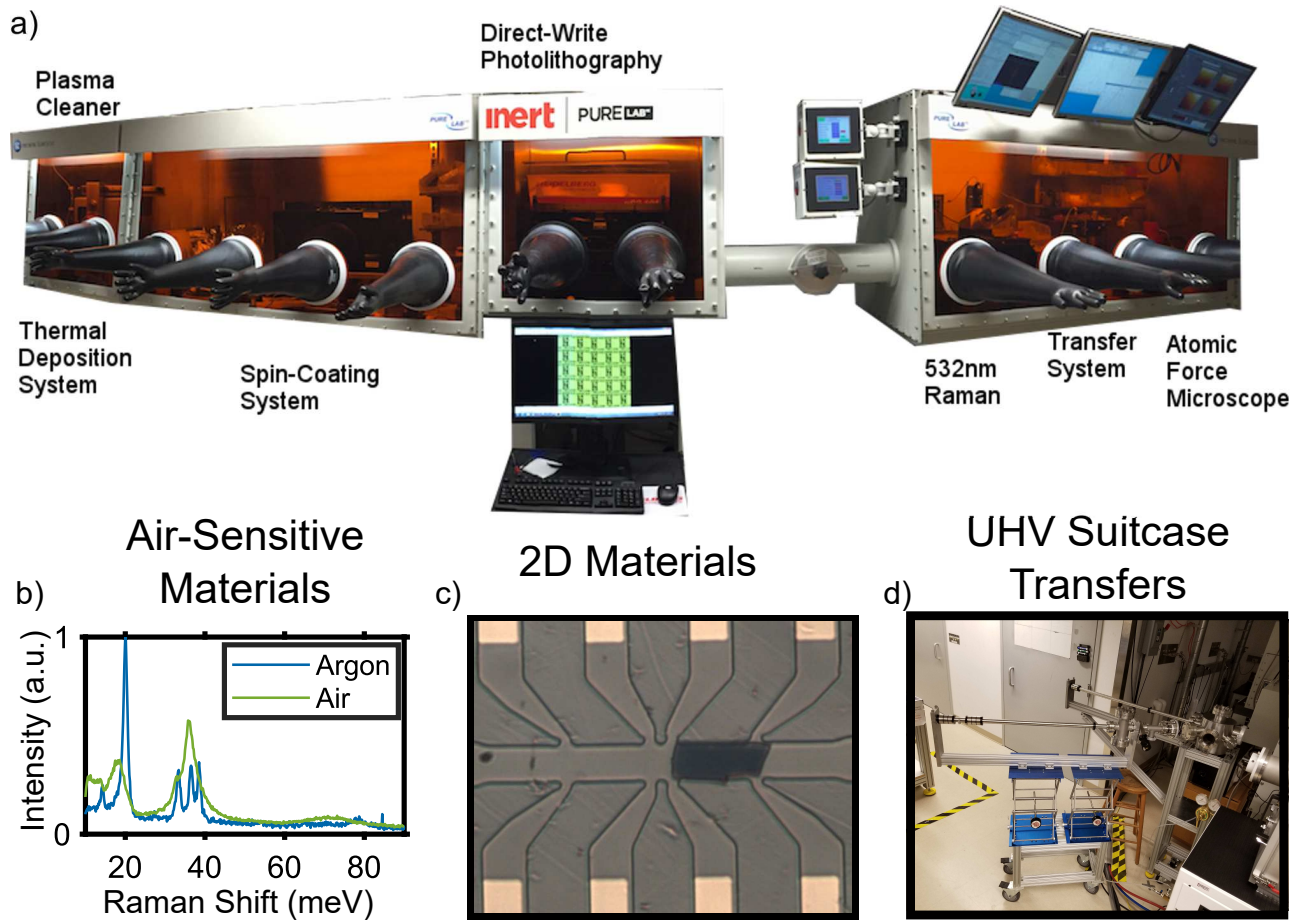}
    \caption{a) Picture of the system. b) Raman spectra measured on $\alpha$-RuCl$_{3}$ showing the difference exfoliation in the inert atmosphere makes. Raman measurements were taken using the WITec Raman System installed in the glovebox. c) Bi$_{2}$Sr$_{2}$CaCu$_{2}$O$_{8+\delta}$ exfoliated onto a thin film of Ga$_{1-x}$Mn$_{x}$As. The film was then etched into a double hall-bar structure around the flake. d) Photo of the UHV suitcase during a device transfer from the glovebox to the low-temperature Raman system. The UHV suitcase is attached to the back of the glovebox.}
    \label{fig:Overview}
\end{figure}

Fabrication of devices at the nano-scale is central to future efforts in exploring novel quantum phases of matter and building next-generation devices. Previously this was achieved by creating dedicated facilities where the entire space is filtered and dust minimized via special air handling and attire for all who enter. While these cleanrooms minimize the amount of dust and other particles that can damage mesoscale devices, they do not protect the samples from either oxygen or water. At the same time they require extremely expensive and energy-intensive investments. In contrast, gloveboxes provide an inert atmosphere for working with oxygen and water sensitive materials, with greatly reduced initial and operational cost.\cite{Chae2016} However, performing nanolithography in a glovebox risks contaminating the rest of the inert environment due to the various solvents involved. With these issues in mind, we have designed and constructed the cleanroom-in-a-glovebox to bridge the gap between these two approaches in order to prepare, fabricate, and characterize various scientific samples entirely within an inert argon atmosphere. The cleanroom-in-a-glovebox contains two separate work chambers, one is devoted entirely to lithography and the other to preparation and characterization (Fig. \ref{fig:Overview}a). The system can be operated with minimal training, no need for special attire (i.e. gowning), and far fewer demands on the building. As such, the described cleanroom in a glovebox produces high-quality devices yet requires far lower initial investment and operational cost than a traditional cleanroom.\cite{Liddle2020} This makes the system described crucial in future efforts at training the quantum workforce and development of novel devices with a wider range of materials.

\begin{figure}[h]
    \centering
    \includegraphics[width = \textwidth]{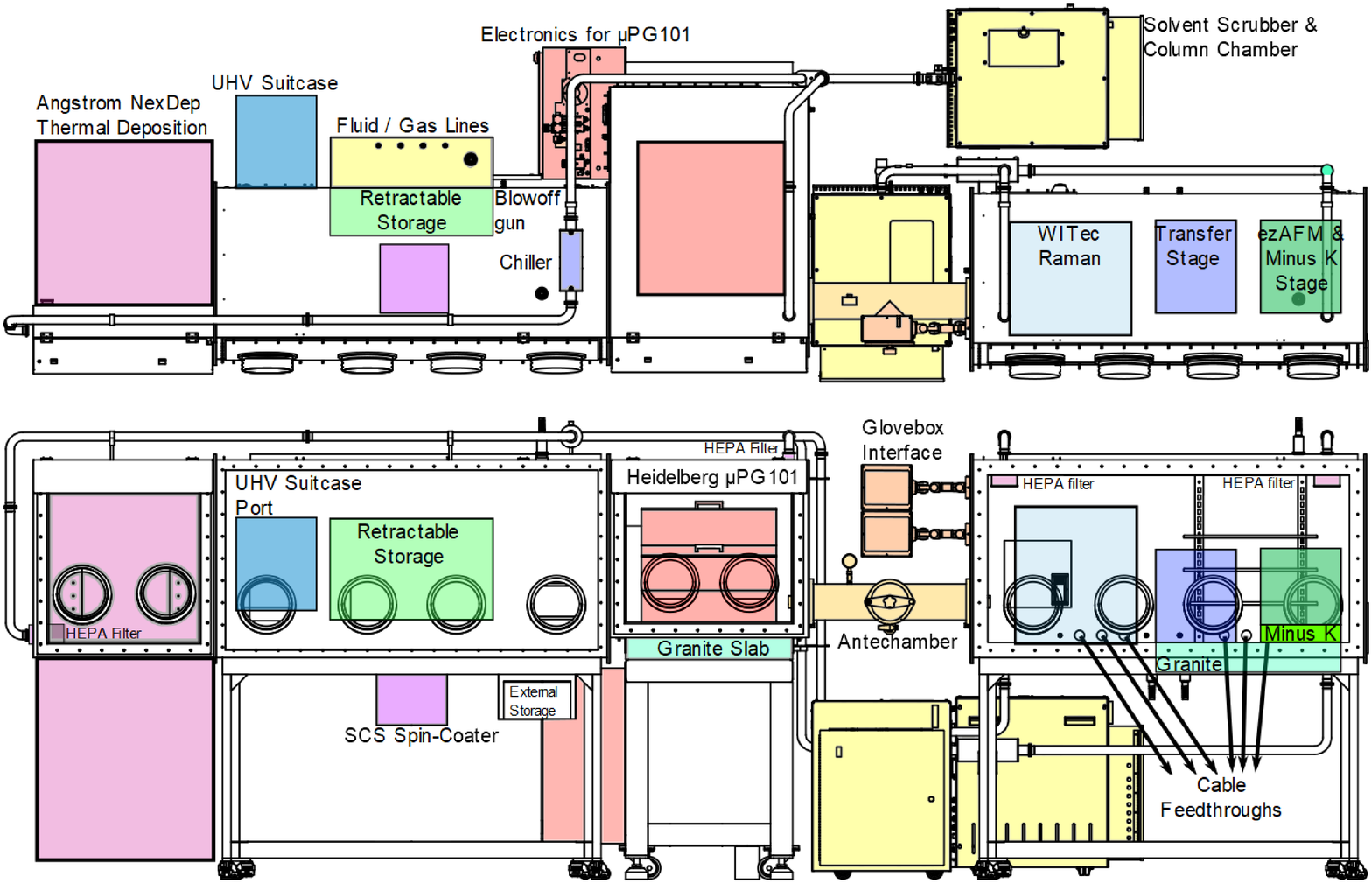}
    \caption{Schematic of the glovebox from both a top-view and front-view.}
    \label{fig:GloveboxSchematic}
\end{figure}

An overview of the system is shown in Fig. \ref{fig:Overview} and a more in-depth schematic of the glovebox is shown in Fig. \ref{fig:GloveboxSchematic}. The lithography chamber, (discussed in the ``Fabrication'' section), contains a Heidelberg $\mu$PG101 Direct-Write system, an Angstrom NexDep Thermal Deposition and Plasma Etching system, and a Spin-Coating Systems G3 Spin Coater. The characterization chamber contains a WITec alpha300R confocal Raman system (Fig. \ref{fig:Overview}b \& Fig. \ref{fig:Characterization}b), a Nanomagnetics ezAFM (Fig. \ref{fig:Characterization}a), a home-built 2D material dry-transfer system, electronic BNC and banana cable feedthroughs. These two chambers are connected via a small antechamber which allows us to transfer samples into and out of the gloveboxes while also enabling simple transfer between boxes without contamination. Lastly, attached to the back of the glovebox is an intermediate chamber for attaching a vacuum suitcase (Fig. \ref{fig:Overview}d). This allows receiving from and transferring to a wide array of ultra-high vacuum (UHV) systems, providing compatibility with electron-beam systems, scanning tunneling microscopy (STM), molecular beam epitaxy (MBE), angle resolved photoemission spectroscopy (ARPES) and other cutting edge tools. Furthermore, the modular nature of the glovebox allows for future equipment to be attached to the glovebox with relative ease. As such our processes and design enable a range of scientific tools on nanoscale, air-sensitive materials, while simultaneously reducing the time, training and cost involved. 

During normal operation, the characterization glovebox maintains O$_{2}$ and  H$_{2}$O levels below the resolution of the sensors (0.1 ppm) while the fabrication box maintains $<$ 1 ppm. This is achieved by continuously running the argon through a copper column chamber which the oxygen and water adhere to, thus removing them from the argon. If the oxygen and water levels are unable to be maintained below 5 ppm the glovebox is purged with argon until the values are back to the normal operating levels. The argon is additionally run through two 300 nm HEPA filters during this recirculation process. When transferring objects into the glovebox, users are required to wipe down the objects with isopropanol wipes to limit the amount of dust that is transferred with the objects. Users are also required to wear nitrile gloves on top of the butyl gloves of the box in order to avoid cross contamination of adhesives. After exfoliation, the nitrile gloves are removed and all associated exfoliation waste is wrapped in the gloves and immediately taken out of the glovebox. As a result of these procedures and the HEPA filters, after four years of continuous operation there were no particles $\geq$ 0.5 $\mu$m measured and an average of 6,800 particles that were $\geq$ 0.3 $\mu$m per cubic meter. This is the equivalent of a class 100 cleanroom according to ISO cleanroom standards.\cite{ISO14644}. We anticipate even better particle levels could be achieved via use of higher quality filters. Furthermore, we did not have to change the HEPA filters over the first four years of operation whereas a typical cleanroom routinely changes the pre-filters every 6 to 12 months.

\section{\label{sec:level2}Materials Characterization}
\begin{figure}[h]
    \centering
    \includegraphics{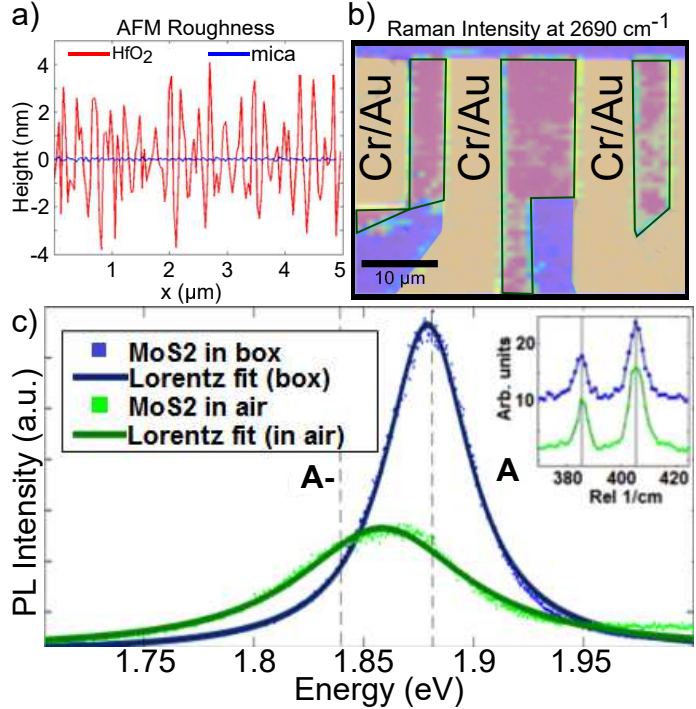}
    \caption{a) Line scan using the \textit{in-situ} AFM on halfnium oxide and mica substrates. This data demonstrates both the effectiveness of our vibration isolation methods and the atomically flat surface of mica. b) Area scan of a patterned CVD graphene device using the \textit{in-situ} Raman system. Graphene is outlined in green while color represents the intensity of the 2D-peak. c) Photoluminescence of MoS$_{2}$ exfoliated on mica. Blue data represents MoS$_{2}$ which was exfoliated onto mica in the glovebox while green represents exfoliation in the ambient environment. The inset shows Raman spectroscopy in the same conditions as the PL. We note that since the phonon modes do not shift in energy we can attribute this drastic change in PL to the inert glovebox environment and not to the dielectric characteristics of the substrate.}
    \label{fig:Characterization}
\end{figure}

Atomic force microscopy (AFM) is an invaluable tool for characterizing materials. In the case of mesoscale physics, AFM is used to discern the thickness of exfoliated 2D materials. In other cases it characterizes the roughness of a substrate (such as in \ref{fig:Characterization}a) or sample. In order to resolve such small features, great care was taken to isolate the AFM system from environmental vibrations. This is more difficult than usual in a glovebox as there are quite a few vibrations that arise from the gas-circulation system. To combat the environmental vibrations the ezAFM and transfer stage were placed on a heavy granite slab. An additional Minus-K Vibration isolation stage was employed for the ezAFM and care was taken to ensure the cables were well secured to each other but did not touch the glovebox directly. The results of this are seen in Fig. \ref{fig:Characterization}a where we took an AFM scan of mica, an atomically flat substrate. The noise levels of the scan are less than 5 $\text{\normalfont{\AA}}$ in magnitude (the resolution of the ezAFM). To ensure rougher features can be resolved, this was compared with the AFM from HfO$_{2}$ film on a Si substrate grown by atomic-layer deposition. We note that ezAFM works with voice coils and thus is substantially less expensive and easier to use than a typical AFM system. Nonetheless, we anticipate a further reduction in noise with more traditional piezo-based scanning probes. 
\par

Raman spectroscopy can reveal the quality, doping level, thickness, symmetry, and cleanliness of samples.\cite{Ferrari:2013jx,Shahil:2010fg,Zhou2018,lei2019high,PhysRevB.82.064503,BTSAPLlocal2016} For example, the ratio of the 2D peak to the G peak in graphene is commonly employed to discern how disordered the sample is.\cite{wu2018raman} With our Raman system's mapping capabilities, we determined the spatial distribution of the disorder after the fabrication of CVD graphene such as in Fig. \ref{fig:Characterization}b. The WITec system also allows us to measure photoluminescence (PL) with a simple switch of energy ranges. PL is a useful measurement technique when working with materials such as MoS$_{2}$, as it quickly identifies single-layer flakes, and provides insight into the interaction with the substrate.\cite{YHlee2017review,yin2011single,Butler:2013ha} We observed another advantage of the glovebox here. Namely, mica is known to have charged potassium ions on the surface after cleaving, but is quickly neutralized in air.\cite{Lui2009} When exfoliating MoS$_{2}$ directly to the mica we found the PL consistent with the mica taking the MoS$_{2}$ from n-type to intrinsic.\cite{Mak2013,Ross2013} (see Fig. \ref{fig:Characterization}c) 

It is crucial to overcome the ``glovebox-specific'' problems to obtain the high-quality Raman and PL data. These are two-fold, first additional light contamination adding unwanted background signals and change in focus or position of the sample due to vibrations, air currents, and temperature fluctuations. To minimize these effects a simple casing was placed around the entire Raman system, using black plastic sheets and 80-20 aluminum bars. Combined with careful isolation of the fibers and wires via foam sealing to the glovebox, the case enabled high-resolution Raman and PL area-scans like the one shown in Fig. \ref{fig:Characterization}b and c.

\section{\label{sec:level3}Sample Fabrication}
The ability to create mesoscopic heterostructures has been crucial in the study of 2D materials, by enabling new physical effects and allowing encapsulation for removal to air.\cite{Dean:2012ht,Sharpe2019,Wu2018WTe2,stepanov2018long,Tang2017WTe2,Zareapour:2012ja,Island2019,Chae2016} However, this relies on minimizing additional contaminants from solvents. Thus we constructed a standard dry-transfer system in the characterization chamber.\cite{Castellanos_Gomez_2014} To ensure excellent alignment and minimal drift during transfer, the stage was placed on a thick granite slab, with the required wires and tubing isolated from touching the glovebox chamber directly. The transfer stage has six, fully-motorized stages, three of which are piezo-based Picomotor stages with a 30 nm step size providing precise positioning of the samples relative to one another, such as the heterostructure shown in Fig. \ref{fig:Overview}c. Furthermore, this system has produced a number of complex devices including the realization of Coulomb blockade into atomic defects in a 2D heterostructure\cite{Brotons-Gisbert2019}, observation of hinge modes in a higher order topological superconductor,\cite{Gray2019} and CVD graphene sensors of bacteria with single cell resolution\cite{KUMAR2020112123}. 

\begin{figure}[h]
    \centering
    \includegraphics{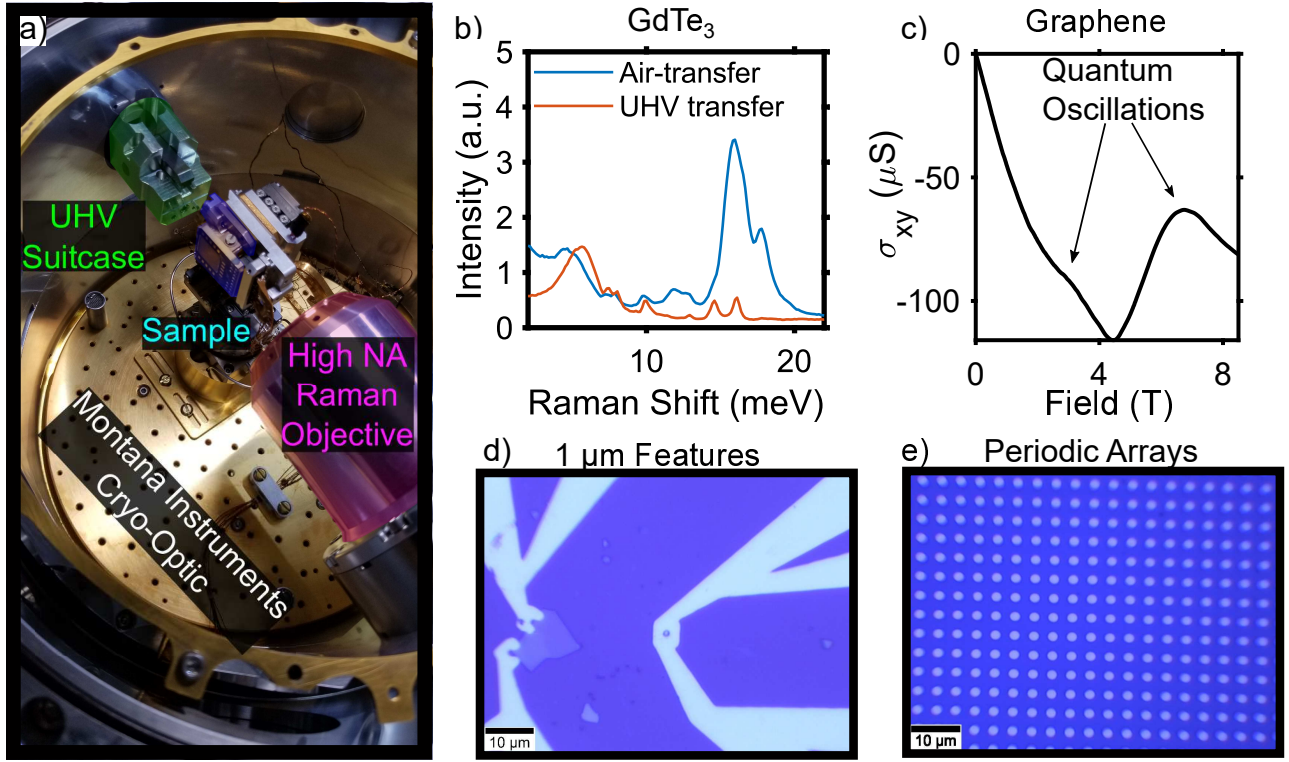}
    \caption{a) Photo taken when transferring a sample from the glovebox into the Low-temperature Raman system. Highlighted in green is the transfer arm from the UHV suitcase, in blue is the sample holder mounted onto the cryocooler, and in purple is the high NA Raman objective. b) Comparison of Raman spectra of GdTe$_{3}$ demonstrating the degradation of the sample when exposed to air for even a few minutes, but is unaffected by first being in the glovebox and then suitcase transferred. c) Hall conductance versus Magnetic Field for a CVD grown graphene sample, fabricated into a hall bar geometry in the glovebox. Even at 7 K, the sample shows quantum oscillations (see arrows). d) Superconducting aluminum loops of 1 $\mu$m radius fabricated onto FeTe$_{0.55}$Se$_{0.45}$ demonstrating the single micron resolution of the $\mu$PG101 photolithography system. e) Periodic arrays of 1 $\mu$m gold pillars.}
    \label{fig:FabricationFigure}
\end{figure}

One of the key features of our cleanroom-in-a-glovebox is our photolithography capabilities. In our fabrication chamber, we have an SCS G3 Spin Coater, Angstrom Engineering NexDep physical vapor deposition system, a UHV suitcase transfer system, and a Heidelberg $\mu$PG101 Direct-Write system. The glovebox column has a solvent scrubber installed, which allows for small amounts of solvent to be removed from the system. This keeps the rest of the environment clean while using the photolithographic, lift-off, and cleaning solvents. We employ the use of Qorpak bottles to limit the exposure of solvents and other liquids to the glovebox atmosphere. These bottles have a PTFE liner in the caps that are resistant to most chemicals while also providing a low moisture transmission rate when sealed. The bottles remain sealed except for the brief times needed. Furthermore, we employ the use of some administrative procedures such as using activated charcoal as a passive solvent absorbent when exposing the environment to liquids and purging the box with argon after fabrication. The result of such efforts are shown in Raman, PL, and AFM scans of materials where long term exposure to the fabrication chamber revealed no evidence for additional contamination. This is further attested to by our ability to observe quantum oscillations at relatively low fields in graphene devices fabricated inside (Fig. \ref{fig:FabricationFigure}c). We note that there are a variety of options for fabrication which completely eliminate water from the process, including shadow (or stencil) mask techniques and transferring flakes onto pre-written contacts.\cite{Zhao2019,Zalic2019}

The lack of contamination along with the alignment abilities of the mask-less system was crucial in creating high-quality devices and periodic structures (see Fig. \ref{fig:FabricationFigure}d and \ref{fig:FabricationFigure}e). The $\mu$PG101 has a resolution of $1~\mu$m with a 20 nm registry, optical auto-focus and can write up to a 5-inch wafer in one run. We note optical auto-focusing is required as the changing dynamics of the glovebox air prevented the use of standard pressure alignment. The $\mu$PG101 stage runs on an air-bearing that is normally supplied with compressed air from the building, but this is not possible while in a glovebox as the unfiltered air would vent directly into the clean environment. Instead, we inserted a T-junction into the argon path from the cylinder where one side of the junction goes into the cylinder to supply the glovebox and the other supplies the stage with argon for the air-bearing. Not only does this solve the air-bearing problem but it also speeds up the removal of excess solvents and water from the clean atmosphere. To shut off the air-bearing when the system is not in use, we installed a cutoff valve after the T-junction that is closed when the stages are not in use.
\par

In a typical nanofabrication process, one must develop and dry the samples in air before moving them into a deposition tool. With an \textit{in-situ} thermal deposition system glovebox users are able to develop and dry the sample in the inert argon environment before transferring them into the deposition tool. Furthermore, the deposition tool contains an \textit{in-situ} plasma-cleaning system so samples can be de-scummed in high vacuum immediately before the deposition of metals. This step can be critical in establishing good electrical contact to certain materials.\cite{Zalic2019} After the deposition, small amounts of aluminum can be evaporated onto the samples followed by exposure to a 0.1\% oxygen environment, creating an air-protection layer of alumina.\cite{Damasco2019tunnel} This layer of alumina can also be used to protect samples against photoresists during nanofabrication processes as it is easily removed by TMAH-based developers. For example, when fabricating CVD graphene devices we first deposit a layer of alumina before spin-coating photoresists while the rest of the fabrication process remains exactly the same, including energy dosage and developing times. The areas of photoresist that are developed out also allow for the developer to come in contact with the alumina, removing it as well. Thus we are still able to make good electrical contact to the graphene while preventing contact with the photoresist and other potential dopants.

The deposition tool also opens to the outside allowing users to clean samples with argon plasma or thermal annealing before loading them into the glovebox. An example is our fabrication of CVD graphene devices for use in bio-sensing applications. The CVD graphene is grown on copper foil and thus must be transferred onto SiO$_{2}$/Si wafers via wet transfer.\cite{doi:10.1021/acsnano.6b04110} In order to clean the graphene, we bake the samples in the deposition tool at 350 $^{o}$C in 10$^{-7}$ mBar pressure for nine hours before alumina deposition (described above) then subsequently transfer samples into the glovebox for patterning. The result of this is samples that are clean enough to not only see quantum oscillations at 8 K and 7 T shown in Fig. \ref{fig:FabricationFigure}c, but are also able to be used as single-bacterium bio-detectors.\cite{KUMAR2020112123}

\section{\label{sec:level4}Ultra High Vacuum Suitcase}
After fabrication, samples typically must be taken out of the glovebox to be measured in more specialized pieces of equipment such as surface-sensitive (STM, APRES) or low-temperature transport and optical probes. Furthermore, many new materials and heterostructures are first created by MBE, requiring \textit{in-situ} probes to determine their device characteristics.\cite{Gerber2017FeSe,2016SuScT..29l3001W,Hellman2017Rev} This presents a chance for the samples to see air and degrade. Typically this is avoided by coating the samples with a ``capping-layer'' (e.g. alumina) or covering mesoscale samples with hBN. However, samples may interact with these materials in unexpected ways such as accidental electrical shorting if the alumina contains many pinholes or if the hBN induces strain into the samples. The addition of hBN to an exfoliated flake could cause additional complexities including changing the dielectric environment or inducing Moir\'e patterns that, while exciting, make reproducibility of devices quite difficult as both layers must be aligned in the same orientation  every time.\cite{Sharpe2019,Woods2014,doi:10.1021/nl5006542,Tran:2019aa,Jin:2019aa,Alexeev:2019aa,Yankowitz2019,Cao2018} Another exciting example of eliminating hBN from air-sensitive devices is the $\beta$-Fe$_{1.1}$Se crystal, where recent experiments have shown enhancements of T$_{c}$ in monolayer films as compared to bulk samples but clean monolayer-devices have yet to be realized.\cite{Gerber2017FeSe,2016SuScT..29l3001W} This is in part due to the air-sensitivity of the system at low layer numbers but is also due to the crystal's sensitivity to strain.\cite{Yang2019} Recent experiments have shown that the T$_{c}$ of $\beta$-Fe$_{1.1}$Se thin films change as much as 10 K with 1\% strain which demonstrates the problem in making hBN encapsulated devices.\cite{Kawai_2018} 

To expand the range of probes and fabrication capabilities of the cleanroom-in-a-glovebox, we designed and built a UHV chamber to couple to various vacuum suitcases (see Fig. \ref{fig:Overview}d). The intermediate chamber has a block for attaching different kinds of sample holders allowing us to transfer materials into the glovebox from MBE and out to STM, low-temperature Raman, or electrical transport systems (e.g. see Fig. \ref{fig:FabricationFigure}a). One measurement system of particular interest is the custom-designed Montana Instruments low-temperature optical cryostat. This system has been described in detail in other works\cite{Tian2016} but has been adapted to be compatible with a UHV suitcase. All of the suitcases follow typical transfer procedures with the addition of a connection to an inlet for argon gas.\cite{Firpo2005} Specifically, after the sample is brought into the intermediate space, the suitcase is valved off and Ar added to bring the chamber to match the glovebox pressure. Once matched the intermediate chamber is opened to the glovebox, where the sample holder is brought in using a second manipulator arm. When transferring devices out of the glovebox a baking step is added to the normal process after vacuuming where the entire chamber is heated to 120$^{o}$C. This step helps remove any excess impurities introduced when exposing the intermediate chamber to the glovebox.

The merits of such work are shown in Fig. \ref{fig:FabricationFigure}b, where we probe the Raman response of GdTe$_{3}$, established to be highly air sensitive.\cite{lei2019high} Two bulk crystals were prepared in the glovebox, then one was transferred into the low-temperature Raman system in air and freshly cleaved just before cool down. The second sample was transferred via the UHV suitcase. The crystal that was transferred in air clearly shows a large tellurium oxide peak around 17 meV that obscures phonon modes.\cite{YaoCGT2Dmat,lei2019high} However, the material transferred via vacuum suitcase revealed sharp phonon modes, with the exception of the CDW amplitude mode at low energies. In addition, we found the Raman response to be much more uniform across the sample surface. 

\section{Conclusions}
Here we demonstrated the construction and operation of a cleanroom-in-a-glovebox. The system combines the inert environment of a glovebox with the fabrication and characterization facilities of a cleanroom. While modifications had to be made to existing equipment and procedures, the result is a fast and efficient fabrication facility that allows devices made from many air-sensitive systems. As a result, we believe our work will motivate future efforts in the development of equipment and techniques in inert atmosphere for next-generation devices. In addition, the far reduced cost, ease of use, and environmental requirements open the door to using this setup in a wider array of educational as well as research settings. 

\section*{Acknowledgments}
The authors are especially grateful to T. Ezell from Witec Instruments, Matt Woods from. Angstrom Instruments Inc., Jake DeGrave and Bobby Riviere for assistance in designing and implementing various components. The authors are grateful to Paige Kelley-Lampen, David Mandrus, and Stephen E. Nagler for providing the RuCl$_{3}$ sample. M.J.G., G.B.O., V.P. and K.S.B acknowledge the primary support of the US Department of Energy (DOE), Office of Science, Office of Basic Energy Sciences under award no. DE-SC0018675. Y.W. acknowledges support from the National Science Foundation, Award No. DMR-1709987, while R.O'C, M.C.D., M.L.R. were supported by grant BIO-1560200.  X.L. acknowledges the support of Boston University. I.Z., B.R. and H.Z. gratefully acknowledges the support from the National Science Foundation Grant No. NSF-DMR-1654041. Work at BNL is supported by the Office of Science, U.S. Department of Energy under Contract No. de-sc0012704. This research is funded in part by the Gordon and Betty Moore Foundation through Grant GBMF9064 to L.M.S.

\section*{Data Availability Statement}
The data that support the findings of this study are available from the corresponding author upon reasonable request.

%\bibliography{CRIAG.bib}

\begin{thebibliography}{46}%
\makeatletter
\providecommand \@ifxundefined [1]{%
 \@ifx{#1\undefined}
}%
\providecommand \@ifnum [1]{%
 \ifnum #1\expandafter \@firstoftwo
 \else \expandafter \@secondoftwo
 \fi
}%
\providecommand \@ifx [1]{%
 \ifx #1\expandafter \@firstoftwo
 \else \expandafter \@secondoftwo
 \fi
}%
\providecommand \natexlab [1]{#1}%
\providecommand \enquote  [1]{``#1''}%
\providecommand \bibnamefont  [1]{#1}%
\providecommand \bibfnamefont [1]{#1}%
\providecommand \citenamefont [1]{#1}%
\providecommand \href@noop [0]{\@secondoftwo}%
\providecommand \href [0]{\begingroup \@sanitize@url \@href}%
\providecommand \@href[1]{\@@startlink{#1}\@@href}%
\providecommand \@@href[1]{\endgroup#1\@@endlink}%
\providecommand \@sanitize@url [0]{\catcode `\\12\catcode `\$12\catcode
  `\&12\catcode `\#12\catcode `\^12\catcode `\_12\catcode `\%12\relax}%
\providecommand \@@startlink[1]{}%
\providecommand \@@endlink[0]{}%
\providecommand \url  [0]{\begingroup\@sanitize@url \@url }%
\providecommand \@url [1]{\endgroup\@href {#1}{\urlprefix }}%
\providecommand \urlprefix  [0]{URL }%
\providecommand \Eprint [0]{\href }%
\providecommand \doibase [0]{http://dx.doi.org/}%
\providecommand \selectlanguage [0]{\@gobble}%
\providecommand \bibinfo  [0]{\@secondoftwo}%
\providecommand \bibfield  [0]{\@secondoftwo}%
\providecommand \translation [1]{[#1]}%
\providecommand \BibitemOpen [0]{}%
\providecommand \bibitemStop [0]{}%
\providecommand \bibitemNoStop [0]{.\EOS\space}%
\providecommand \EOS [0]{\spacefactor3000\relax}%
\providecommand \BibitemShut  [1]{\csname bibitem#1\endcsname}%
\let\auto@bib@innerbib\@empty
%</preamble>
\bibitem [{\citenamefont {Chae}\ \emph {et~al.}(2016)\citenamefont {Chae},
  \citenamefont {Jin}, \citenamefont {Kim}, \citenamefont {Chung},
  \citenamefont {Na}, \citenamefont {Nam}, \citenamefont {Kim}, \citenamefont
  {Perello}, \citenamefont {Jeong}, \citenamefont {Ly},\ and\ \citenamefont
  {Lee}}]{Chae2016}%
  \BibitemOpen
  \bibfield  {author} {\bibinfo {author} {\bibfnamefont {S.~H.}\ \bibnamefont
  {Chae}}, \bibinfo {author} {\bibfnamefont {Y.}~\bibnamefont {Jin}}, \bibinfo
  {author} {\bibfnamefont {T.~S.}\ \bibnamefont {Kim}}, \bibinfo {author}
  {\bibfnamefont {D.~S.}\ \bibnamefont {Chung}}, \bibinfo {author}
  {\bibfnamefont {H.}~\bibnamefont {Na}}, \bibinfo {author} {\bibfnamefont
  {H.}~\bibnamefont {Nam}}, \bibinfo {author} {\bibfnamefont {H.}~\bibnamefont
  {Kim}}, \bibinfo {author} {\bibfnamefont {D.~J.}\ \bibnamefont {Perello}},
  \bibinfo {author} {\bibfnamefont {H.~Y.}\ \bibnamefont {Jeong}}, \bibinfo
  {author} {\bibfnamefont {T.~H.}\ \bibnamefont {Ly}}, \ and\ \bibinfo {author}
  {\bibfnamefont {Y.~H.}\ \bibnamefont {Lee}},\ }\href {\doibase
  10.1021/acsnano.5b06680} {\bibfield  {journal} {\bibinfo  {journal} {ACS
  Nano}\ }\textbf {\bibinfo {volume} {10}},\ \bibinfo {pages} {1309} (\bibinfo
  {year} {2016})}\BibitemShut {NoStop}%
\bibitem [{\citenamefont {Liddle}\ \emph {et~al.}(2020)\citenamefont {Liddle},
  \citenamefont {Bowser}, \citenamefont {Ilic},\ and\ \citenamefont
  {Luciani}}]{Liddle2020}%
  \BibitemOpen
  \bibfield  {author} {\bibinfo {author} {\bibfnamefont {J.~A.}\ \bibnamefont
  {Liddle}}, \bibinfo {author} {\bibfnamefont {J.}~\bibnamefont {Bowser}},
  \bibinfo {author} {\bibfnamefont {B.~R.}\ \bibnamefont {Ilic}}, \ and\
  \bibinfo {author} {\bibfnamefont {V.}~\bibnamefont {Luciani}},\ }\href
  {\doibase 10.6028/jres.125.009} {\bibfield  {journal} {\bibinfo  {journal} {J
  Res Natl Inst Stan}\ }\textbf {\bibinfo {volume} {125}} (\bibinfo {year}
  {2020})}\BibitemShut {NoStop}%
\bibitem [{ISO 114644-1:2015(2015)}]{ISO14644}%
  \BibitemOpen
  \bibfield  {author} {ISO 114644-1:2015,\ }\href@noop {} {\enquote {\bibinfo
  {title} {{Cleanrooms and associated controlled environments — Part 1:
  Classification of air cleanliness by particle concentration}},}\ }\bibinfo
  {type} {Standard}\ (\bibinfo  {institution} {International Organization for
  Standardization},\ \bibinfo {address} {Geneva, CH},\ \bibinfo {year}
  {2015})\BibitemShut {NoStop}%
\bibitem [{\citenamefont {Ferrari}\ and\ \citenamefont
  {Basko}(2013)}]{Ferrari:2013jx}%
  \BibitemOpen
  \bibfield  {author} {\bibinfo {author} {\bibfnamefont {A.~C.}\ \bibnamefont
  {Ferrari}}\ and\ \bibinfo {author} {\bibfnamefont {D.~M.}\ \bibnamefont
  {Basko}},\ }\href@noop {} {\bibfield  {journal} {\bibinfo  {journal} {Nature
  Nanotechnology}\ }\textbf {\bibinfo {volume} {8}},\ \bibinfo {pages} {235}
  (\bibinfo {year} {2013})}\BibitemShut {NoStop}%
\bibitem [{\citenamefont {Shahil}\ \emph {et~al.}(2010)\citenamefont {Shahil},
  \citenamefont {Hossain}, \citenamefont {Teweldebrhan},\ and\ \citenamefont
  {Balandin}}]{Shahil:2010fg}%
  \BibitemOpen
  \bibfield  {author} {\bibinfo {author} {\bibfnamefont {K.~M.~F.}\
  \bibnamefont {Shahil}}, \bibinfo {author} {\bibfnamefont {M.~Z.}\
  \bibnamefont {Hossain}}, \bibinfo {author} {\bibfnamefont {D.}~\bibnamefont
  {Teweldebrhan}}, \ and\ \bibinfo {author} {\bibfnamefont {A.~A.}\
  \bibnamefont {Balandin}},\ }\href@noop {} {\bibfield  {journal} {\bibinfo
  {journal} {Applied Physics Letters}\ }\textbf {\bibinfo {volume} {96}},\
  \bibinfo {pages} {3103} (\bibinfo {year} {2010})}\BibitemShut {NoStop}%
\bibitem [{\citenamefont {Zhou}\ \emph {et~al.}(2019)\citenamefont {Zhou},
  \citenamefont {Wang}, \citenamefont {Osterhoudt}, \citenamefont
  {Lampen-Kelley}, \citenamefont {Mandrus}, \citenamefont {He}, \citenamefont
  {Burch},\ and\ \citenamefont {Henriksen}}]{Zhou2018}%
  \BibitemOpen
  \bibfield  {author} {\bibinfo {author} {\bibfnamefont {B.}~\bibnamefont
  {Zhou}}, \bibinfo {author} {\bibfnamefont {Y.}~\bibnamefont {Wang}}, \bibinfo
  {author} {\bibfnamefont {G.~B.}\ \bibnamefont {Osterhoudt}}, \bibinfo
  {author} {\bibfnamefont {P.}~\bibnamefont {Lampen-Kelley}}, \bibinfo {author}
  {\bibfnamefont {D.}~\bibnamefont {Mandrus}}, \bibinfo {author} {\bibfnamefont
  {R.}~\bibnamefont {He}}, \bibinfo {author} {\bibfnamefont {K.~S.}\
  \bibnamefont {Burch}}, \ and\ \bibinfo {author} {\bibfnamefont {E.~A.}\
  \bibnamefont {Henriksen}},\ }\href {\doibase
  https://doi.org/10.1016/j.jpcs.2018.01.026} {\bibfield  {journal} {\bibinfo
  {journal} {Journal of Physics and Chemistry of Solids}\ }\textbf {\bibinfo
  {volume} {128}},\ \bibinfo {pages} {291} (\bibinfo {year} {2019})}\BibitemShut {NoStop}%
\bibitem [{\citenamefont {Lei}\ \emph {et~al.}(2020)\citenamefont {Lei},
  \citenamefont {Lin}, \citenamefont {Jia}, \citenamefont {Gray}, \citenamefont
  {Topp}, \citenamefont {Farahi}, \citenamefont {Klemenz}, \citenamefont {Gao},
  \citenamefont {Rodolakis}, \citenamefont {McChesney}, \citenamefont {Ast},
  \citenamefont {Yazdani}, \citenamefont {Burch}, \citenamefont {Wu},
  \citenamefont {Ong},\ and\ \citenamefont {Schoop}}]{lei2019high}%
  \BibitemOpen
  \bibfield  {author} {\bibinfo {author} {\bibfnamefont {S.}~\bibnamefont
  {Lei}}, \bibinfo {author} {\bibfnamefont {J.}~\bibnamefont {Lin}}, \bibinfo
  {author} {\bibfnamefont {Y.}~\bibnamefont {Jia}}, \bibinfo {author}
  {\bibfnamefont {M.}~\bibnamefont {Gray}}, \bibinfo {author} {\bibfnamefont
  {A.}~\bibnamefont {Topp}}, \bibinfo {author} {\bibfnamefont {G.}~\bibnamefont
  {Farahi}}, \bibinfo {author} {\bibfnamefont {S.}~\bibnamefont {Klemenz}},
  \bibinfo {author} {\bibfnamefont {T.}~\bibnamefont {Gao}}, \bibinfo {author}
  {\bibfnamefont {F.}~\bibnamefont {Rodolakis}}, \bibinfo {author}
  {\bibfnamefont {J.~L.}\ \bibnamefont {McChesney}}, \bibinfo {author}
  {\bibfnamefont {C.~R.}\ \bibnamefont {Ast}}, \bibinfo {author} {\bibfnamefont
  {A.}~\bibnamefont {Yazdani}}, \bibinfo {author} {\bibfnamefont {K.~S.}\
  \bibnamefont {Burch}}, \bibinfo {author} {\bibfnamefont {S.}~\bibnamefont
  {Wu}}, \bibinfo {author} {\bibfnamefont {N.~P.}\ \bibnamefont {Ong}}, \ and\
  \bibinfo {author} {\bibfnamefont {L.~M.}\ \bibnamefont {Schoop}},\ }\href
  {\doibase 10.1126/sciadv.aay6407} {\bibfield  {journal} {\bibinfo  {journal}
  {Science Advances}\ }\textbf {\bibinfo {volume} {6}} (\bibinfo {year}
  {2020})}\BibitemShut {NoStop}%
\bibitem [{\citenamefont {Sandilands}\ \emph {et~al.}(2010)\citenamefont
  {Sandilands}, \citenamefont {Shen}, \citenamefont {Chugunov}, \citenamefont
  {Zhao}, \citenamefont {Ono}, \citenamefont {Ando},\ and\ \citenamefont
  {Burch}}]{PhysRevB.82.064503}%
  \BibitemOpen
  \bibfield  {author} {\bibinfo {author} {\bibfnamefont {L.~J.}\ \bibnamefont
  {Sandilands}}, \bibinfo {author} {\bibfnamefont {J.~X.}\ \bibnamefont
  {Shen}}, \bibinfo {author} {\bibfnamefont {G.~M.}\ \bibnamefont {Chugunov}},
  \bibinfo {author} {\bibfnamefont {S.~Y.~F.}\ \bibnamefont {Zhao}}, \bibinfo
  {author} {\bibfnamefont {S.}~\bibnamefont {Ono}}, \bibinfo {author}
  {\bibfnamefont {Y.}~\bibnamefont {Ando}}, \ and\ \bibinfo {author}
  {\bibfnamefont {K.~S.}\ \bibnamefont {Burch}},\ }\href@noop {} {\bibfield
  {journal} {\bibinfo  {journal} {Phys. Rev. B}\ }\textbf {\bibinfo {volume}
  {82}},\ \bibinfo {pages} {064503} (\bibinfo {year} {2010})}\BibitemShut
  {NoStop}%
\bibitem [{\citenamefont {Tian}\ \emph
  {et~al.}(2016{\natexlab{a}})\citenamefont {Tian}, \citenamefont {Osterhoudt},
  \citenamefont {Jia}, \citenamefont {Cava},\ and\ \citenamefont
  {Burch}}]{BTSAPLlocal2016}%
  \BibitemOpen
  \bibfield  {author} {\bibinfo {author} {\bibfnamefont {Y.}~\bibnamefont
  {Tian}}, \bibinfo {author} {\bibfnamefont {G.~B.}\ \bibnamefont
  {Osterhoudt}}, \bibinfo {author} {\bibfnamefont {S.}~\bibnamefont {Jia}},
  \bibinfo {author} {\bibfnamefont {R.~J.}\ \bibnamefont {Cava}}, \ and\
  \bibinfo {author} {\bibfnamefont {K.~S.}\ \bibnamefont {Burch}},\ }\href
  {\doibase http://dx.doi.org/10.1063/1.4941022} {\bibfield  {journal}
  {\bibinfo  {journal} {Appl. Phys. Lett.}\ }\textbf {\bibinfo {volume}
  {108}},\ \bibinfo {pages} {041911} (\bibinfo {year}
  {2016}{\natexlab{a}})}\BibitemShut {NoStop}%
\bibitem [{\citenamefont {Wu}\ \emph {et~al.}(2018{\natexlab{a}})\citenamefont
  {Wu}, \citenamefont {Lin}, \citenamefont {Cong}, \citenamefont {Liu},\ and\
  \citenamefont {Tan}}]{wu2018raman}%
  \BibitemOpen
  \bibfield  {author} {\bibinfo {author} {\bibfnamefont {J.-B.}\ \bibnamefont
  {Wu}}, \bibinfo {author} {\bibfnamefont {M.-L.}\ \bibnamefont {Lin}},
  \bibinfo {author} {\bibfnamefont {X.}~\bibnamefont {Cong}}, \bibinfo {author}
  {\bibfnamefont {H.-N.}\ \bibnamefont {Liu}}, \ and\ \bibinfo {author}
  {\bibfnamefont {P.-H.}\ \bibnamefont {Tan}},\ }\href@noop {} {\bibfield
  {journal} {\bibinfo  {journal} {Chemical Society Reviews}\ }\textbf {\bibinfo
  {volume} {47}},\ \bibinfo {pages} {1822} (\bibinfo {year}
  {2018}{\natexlab{a}})}\BibitemShut {NoStop}%
\bibitem [{\citenamefont {Yin}\ \emph {et~al.}(2011)\citenamefont {Yin},
  \citenamefont {Li}, \citenamefont {Li}, \citenamefont {Jiang}, \citenamefont
  {Shi}, \citenamefont {Sun}, \citenamefont {Lu}, \citenamefont {Zhang},
  \citenamefont {Chen},\ and\ \citenamefont {Zhang}}]{yin2011single}%
  \BibitemOpen
  \bibfield  {author} {\bibinfo {author} {\bibfnamefont {Z.}~\bibnamefont
  {Yin}}, \bibinfo {author} {\bibfnamefont {H.}~\bibnamefont {Li}}, \bibinfo
  {author} {\bibfnamefont {H.}~\bibnamefont {Li}}, \bibinfo {author}
  {\bibfnamefont {L.}~\bibnamefont {Jiang}}, \bibinfo {author} {\bibfnamefont
  {Y.}~\bibnamefont {Shi}}, \bibinfo {author} {\bibfnamefont {Y.}~\bibnamefont
  {Sun}}, \bibinfo {author} {\bibfnamefont {G.}~\bibnamefont {Lu}}, \bibinfo
  {author} {\bibfnamefont {Q.}~\bibnamefont {Zhang}}, \bibinfo {author}
  {\bibfnamefont {X.}~\bibnamefont {Chen}}, \ and\ \bibinfo {author}
  {\bibfnamefont {H.}~\bibnamefont {Zhang}},\ }\href@noop {} {\bibfield
  {journal} {\bibinfo  {journal} {ACS nano}\ }\textbf {\bibinfo {volume} {6}},\
  \bibinfo {pages} {74} (\bibinfo {year} {2011})}\BibitemShut {NoStop}%
\bibitem [{\citenamefont {Butler}\ \emph {et~al.}(2013)\citenamefont {Butler},
  \citenamefont {Hollen}, \citenamefont {Cao}, \citenamefont {Cui},
  \citenamefont {Gupta}, \citenamefont {Guti{\'e}rrez}, \citenamefont {Heinz},
  \citenamefont {Hong}, \citenamefont {Huang}, \citenamefont {Ismach},
  \citenamefont {Johnston-Halperin}, \citenamefont {Kuno}, \citenamefont
  {Plashnitsa}, \citenamefont {Robinson}, \citenamefont {Ruoff}, \citenamefont
  {Salahuddin}, \citenamefont {Shan}, \citenamefont {Shi}, \citenamefont
  {Spencer}, \citenamefont {Terrones}, \citenamefont {Windl},\ and\
  \citenamefont {Goldberger}}]{Butler:2013ha}%
  \BibitemOpen
  \bibfield  {author} {\bibinfo {author} {\bibfnamefont {S.~Z.}\ \bibnamefont
  {Butler}}, \bibinfo {author} {\bibfnamefont {S.~M.}\ \bibnamefont {Hollen}},
  \bibinfo {author} {\bibfnamefont {L.}~\bibnamefont {Cao}}, \bibinfo {author}
  {\bibfnamefont {Y.}~\bibnamefont {Cui}}, \bibinfo {author} {\bibfnamefont
  {J.~A.}\ \bibnamefont {Gupta}}, \bibinfo {author} {\bibfnamefont {H.~R.}\
  \bibnamefont {Guti{\'e}rrez}}, \bibinfo {author} {\bibfnamefont {T.~F.}\
  \bibnamefont {Heinz}}, \bibinfo {author} {\bibfnamefont {S.~S.}\ \bibnamefont
  {Hong}}, \bibinfo {author} {\bibfnamefont {J.}~\bibnamefont {Huang}},
  \bibinfo {author} {\bibfnamefont {A.~F.}\ \bibnamefont {Ismach}}, \bibinfo
  {author} {\bibfnamefont {E.}~\bibnamefont {Johnston-Halperin}}, \bibinfo
  {author} {\bibfnamefont {M.}~\bibnamefont {Kuno}}, \bibinfo {author}
  {\bibfnamefont {V.~V.}\ \bibnamefont {Plashnitsa}}, \bibinfo {author}
  {\bibfnamefont {R.~D.}\ \bibnamefont {Robinson}}, \bibinfo {author}
  {\bibfnamefont {R.~S.}\ \bibnamefont {Ruoff}}, \bibinfo {author}
  {\bibfnamefont {S.}~\bibnamefont {Salahuddin}}, \bibinfo {author}
  {\bibfnamefont {J.}~\bibnamefont {Shan}}, \bibinfo {author} {\bibfnamefont
  {L.}~\bibnamefont {Shi}}, \bibinfo {author} {\bibfnamefont {M.~G.}\
  \bibnamefont {Spencer}}, \bibinfo {author} {\bibfnamefont {M.}~\bibnamefont
  {Terrones}}, \bibinfo {author} {\bibfnamefont {W.}~\bibnamefont {Windl}}, \
  and\ \bibinfo {author} {\bibfnamefont {J.~E.}\ \bibnamefont {Goldberger}},\
  }\href@noop {} {\bibfield  {journal} {\bibinfo  {journal} {ACS Nano}\
  }\textbf {\bibinfo {volume} {7}},\ \bibinfo {pages} {2898} (\bibinfo {year}
  {2013})}\BibitemShut {NoStop}%
\bibitem [{\citenamefont {Lui}\ \emph {et~al.}(2009)\citenamefont {Lui},
  \citenamefont {Liu}, \citenamefont {Mak}, \citenamefont {Flynn},\ and\
  \citenamefont {Heinz}}]{Lui2009}%
  \BibitemOpen
  \bibfield  {author} {\bibinfo {author} {\bibfnamefont {C.~H.}\ \bibnamefont
  {Lui}}, \bibinfo {author} {\bibfnamefont {L.}~\bibnamefont {Liu}}, \bibinfo
  {author} {\bibfnamefont {K.~F.}\ \bibnamefont {Mak}}, \bibinfo {author}
  {\bibfnamefont {G.~W.}\ \bibnamefont {Flynn}}, \ and\ \bibinfo {author}
  {\bibfnamefont {T.~F.}\ \bibnamefont {Heinz}},\ }\href {\doibase
  10.1038/nature08569} {\bibfield  {journal} {\bibinfo  {journal} {Nature}\
  }\textbf {\bibinfo {volume} {462}},\ \bibinfo {pages} {339} (\bibinfo {year}
  {2009})}\BibitemShut {NoStop}%
\bibitem [{\citenamefont {Mak}\ \emph {et~al.}(2013)\citenamefont {Mak},
  \citenamefont {He}, \citenamefont {Lee}, \citenamefont {Lee}, \citenamefont
  {Hone}, \citenamefont {Heinz},\ and\ \citenamefont {Shan}}]{Mak2013}%
  \BibitemOpen
  \bibfield  {author} {\bibinfo {author} {\bibfnamefont {K.~F.}\ \bibnamefont
  {Mak}}, \bibinfo {author} {\bibfnamefont {K.}~\bibnamefont {He}}, \bibinfo
  {author} {\bibfnamefont {C.}~\bibnamefont {Lee}}, \bibinfo {author}
  {\bibfnamefont {G.~H.}\ \bibnamefont {Lee}}, \bibinfo {author} {\bibfnamefont
  {J.}~\bibnamefont {Hone}}, \bibinfo {author} {\bibfnamefont {T.~F.}\
  \bibnamefont {Heinz}}, \ and\ \bibinfo {author} {\bibfnamefont
  {J.}~\bibnamefont {Shan}},\ }\href {\doibase 10.1038/nmat3505} {\bibfield
  {journal} {\bibinfo  {journal} {Nature materials}\ }\textbf {\bibinfo
  {volume} {12}},\ \bibinfo {pages} {207} (\bibinfo {year} {2013})}\BibitemShut
  {NoStop}%
\bibitem [{\citenamefont {Ross}\ \emph {et~al.}(2013)\citenamefont {Ross},
  \citenamefont {Wu}, \citenamefont {Yu}, \citenamefont {Ghimire},
  \citenamefont {Jones}, \citenamefont {Aivazian}, \citenamefont {Yan},
  \citenamefont {Mandrus}, \citenamefont {Xiao}, \citenamefont {Yao},\ and\
  \citenamefont {Xu}}]{Ross2013}%
  \BibitemOpen
  \bibfield  {author} {\bibinfo {author} {\bibfnamefont {J.~S.}\ \bibnamefont
  {Ross}}, \bibinfo {author} {\bibfnamefont {S.}~\bibnamefont {Wu}}, \bibinfo
  {author} {\bibfnamefont {H.}~\bibnamefont {Yu}}, \bibinfo {author}
  {\bibfnamefont {N.~J.}\ \bibnamefont {Ghimire}}, \bibinfo {author}
  {\bibfnamefont {A.~M.}\ \bibnamefont {Jones}}, \bibinfo {author}
  {\bibfnamefont {G.}~\bibnamefont {Aivazian}}, \bibinfo {author}
  {\bibfnamefont {J.}~\bibnamefont {Yan}}, \bibinfo {author} {\bibfnamefont
  {D.~G.}\ \bibnamefont {Mandrus}}, \bibinfo {author} {\bibfnamefont
  {D.}~\bibnamefont {Xiao}}, \bibinfo {author} {\bibfnamefont {W.}~\bibnamefont
  {Yao}}, \ and\ \bibinfo {author} {\bibfnamefont {X.}~\bibnamefont {Xu}},\
  }\href {\doibase 10.1038/ncomms2498} {\bibfield  {journal} {\bibinfo
  {journal} {Nature Communications}\ }\textbf {\bibinfo {volume} {4}},\
  \bibinfo {pages} {1474} (\bibinfo {year} {2013})}\BibitemShut {NoStop}%
\bibitem [{\citenamefont {Sharpe}\ \emph {et~al.}(2019)\citenamefont {Sharpe},
  \citenamefont {Fox}, \citenamefont {Barnard}, \citenamefont {Finney},
  \citenamefont {Watanabe}, \citenamefont {Taniguchi}, \citenamefont
  {Kastner},\ and\ \citenamefont {Goldhaber-Gordon}}]{Sharpe2019}%
  \BibitemOpen
  \bibfield  {author} {\bibinfo {author} {\bibfnamefont {A.~L.}\ \bibnamefont
  {Sharpe}}, \bibinfo {author} {\bibfnamefont {E.~J.}\ \bibnamefont {Fox}},
  \bibinfo {author} {\bibfnamefont {A.~W.}\ \bibnamefont {Barnard}}, \bibinfo
  {author} {\bibfnamefont {J.}~\bibnamefont {Finney}}, \bibinfo {author}
  {\bibfnamefont {K.}~\bibnamefont {Watanabe}}, \bibinfo {author}
  {\bibfnamefont {T.}~\bibnamefont {Taniguchi}}, \bibinfo {author}
  {\bibfnamefont {M.~A.}\ \bibnamefont {Kastner}}, \ and\ \bibinfo {author}
  {\bibfnamefont {D.}~\bibnamefont {Goldhaber-Gordon}},\ }\href {\doibase
  10.1126/science.aaw3780} {\bibfield  {journal} {\bibinfo  {journal}
  {Science}\ }\textbf {\bibinfo {volume} {365}},\ \bibinfo {pages} {605}
  (\bibinfo {year} {2019})}\BibitemShut  {NoStop}%
\bibitem [{\citenamefont {Wu}\ \emph {et~al.}(2018{\natexlab{b}})\citenamefont
  {Wu}, \citenamefont {Fatemi}, \citenamefont {Gibson}, \citenamefont
  {Watanabe}, \citenamefont {Taniguchi}, \citenamefont {Cava},\ and\
  \citenamefont {Jarillo-Herrero}}]{Wu2018WTe2}%
  \BibitemOpen
  \bibfield  {author} {\bibinfo {author} {\bibfnamefont {S.}~\bibnamefont
  {Wu}}, \bibinfo {author} {\bibfnamefont {V.}~\bibnamefont {Fatemi}}, \bibinfo
  {author} {\bibfnamefont {Q.~D.}\ \bibnamefont {Gibson}}, \bibinfo {author}
  {\bibfnamefont {K.}~\bibnamefont {Watanabe}}, \bibinfo {author}
  {\bibfnamefont {T.}~\bibnamefont {Taniguchi}}, \bibinfo {author}
  {\bibfnamefont {R.~J.}\ \bibnamefont {Cava}}, \ and\ \bibinfo {author}
  {\bibfnamefont {P.}~\bibnamefont {Jarillo-Herrero}},\ }\href {\doibase
  10.1126/science.aan6003} {\bibfield  {journal} {\bibinfo  {journal}
  {Science}\ }\textbf {\bibinfo {volume} {359}},\ \bibinfo {pages} {76}
  (\bibinfo {year} {2018}{\natexlab{b}})}\BibitemShut {NoStop}%
\bibitem [{\citenamefont {Stepanov}\ \emph {et~al.}(2018)\citenamefont
  {Stepanov}, \citenamefont {Che}, \citenamefont {Shcherbakov}, \citenamefont
  {Yang}, \citenamefont {Chen}, \citenamefont {Thilahar}, \citenamefont
  {Voigt}, \citenamefont {Bockrath}, \citenamefont {Smirnov}, \citenamefont
  {Watanabe}, \citenamefont {Taniguchi}, \citenamefont {Lake}, \citenamefont
  {Barlas}, \citenamefont {MacDonald},\ and\ \citenamefont
  {Lau}}]{stepanov2018long}%
  \BibitemOpen
  \bibfield  {author} {\bibinfo {author} {\bibfnamefont {P.}~\bibnamefont
  {Stepanov}}, \bibinfo {author} {\bibfnamefont {S.}~\bibnamefont {Che}},
  \bibinfo {author} {\bibfnamefont {D.}~\bibnamefont {Shcherbakov}}, \bibinfo
  {author} {\bibfnamefont {J.}~\bibnamefont {Yang}}, \bibinfo {author}
  {\bibfnamefont {R.}~\bibnamefont {Chen}}, \bibinfo {author} {\bibfnamefont
  {K.}~\bibnamefont {Thilahar}}, \bibinfo {author} {\bibfnamefont
  {G.}~\bibnamefont {Voigt}}, \bibinfo {author} {\bibfnamefont {M.~W.}\
  \bibnamefont {Bockrath}}, \bibinfo {author} {\bibfnamefont {D.}~\bibnamefont
  {Smirnov}}, \bibinfo {author} {\bibfnamefont {K.}~\bibnamefont {Watanabe}},
  \bibinfo {author} {\bibfnamefont {T.}~\bibnamefont {Taniguchi}}, \bibinfo
  {author} {\bibfnamefont {R.~K.}\ \bibnamefont {Lake}}, \bibinfo {author}
  {\bibfnamefont {Y.}~\bibnamefont {Barlas}}, \bibinfo {author} {\bibfnamefont
  {A.~H.}\ \bibnamefont {MacDonald}}, \ and\ \bibinfo {author} {\bibfnamefont
  {C.~N.}\ \bibnamefont {Lau}},\ }\href@noop {} {\bibfield  {journal} {\bibinfo
   {journal} {Nature Physics}\ }\textbf {\bibinfo {volume} {14}},\ \bibinfo
  {pages} {907} (\bibinfo {year} {2018})}\BibitemShut {NoStop}%
\bibitem [{\citenamefont {Tang}\ \emph {et~al.}(2017)\citenamefont {Tang},
  \citenamefont {Zhang}, \citenamefont {Wong}, \citenamefont {Pedramrazi},
  \citenamefont {Tsai}, \citenamefont {Jia}, \citenamefont {Moritz},
  \citenamefont {Claassen}, \citenamefont {Ryu}, \citenamefont {Kahn},
  \citenamefont {Jiang}, \citenamefont {Yan}, \citenamefont {Hashimoto},
  \citenamefont {Lu}, \citenamefont {Moore}, \citenamefont {Hwang},
  \citenamefont {Hwang}, \citenamefont {Hussain}, \citenamefont {Chen},
  \citenamefont {Ugeda}, \citenamefont {Liu}, \citenamefont {Xie},
  \citenamefont {Devereaux}, \citenamefont {Crommie}, \citenamefont {Mo},\ and\
  \citenamefont {Shen}}]{Tang2017WTe2}%
  \BibitemOpen
  \bibfield  {author} {\bibinfo {author} {\bibfnamefont {S.}~\bibnamefont
  {Tang}}, \bibinfo {author} {\bibfnamefont {C.}~\bibnamefont {Zhang}},
  \bibinfo {author} {\bibfnamefont {D.}~\bibnamefont {Wong}}, \bibinfo {author}
  {\bibfnamefont {Z.}~\bibnamefont {Pedramrazi}}, \bibinfo {author}
  {\bibfnamefont {H.-Z.}\ \bibnamefont {Tsai}}, \bibinfo {author}
  {\bibfnamefont {C.}~\bibnamefont {Jia}}, \bibinfo {author} {\bibfnamefont
  {B.}~\bibnamefont {Moritz}}, \bibinfo {author} {\bibfnamefont
  {M.}~\bibnamefont {Claassen}}, \bibinfo {author} {\bibfnamefont
  {H.}~\bibnamefont {Ryu}}, \bibinfo {author} {\bibfnamefont {S.}~\bibnamefont
  {Kahn}}, \bibinfo {author} {\bibfnamefont {J.}~\bibnamefont {Jiang}},
  \bibinfo {author} {\bibfnamefont {H.}~\bibnamefont {Yan}}, \bibinfo {author}
  {\bibfnamefont {M.}~\bibnamefont {Hashimoto}}, \bibinfo {author}
  {\bibfnamefont {D.}~\bibnamefont {Lu}}, \bibinfo {author} {\bibfnamefont
  {R.~G.}\ \bibnamefont {Moore}}, \bibinfo {author} {\bibfnamefont {C.-C.}\
  \bibnamefont {Hwang}}, \bibinfo {author} {\bibfnamefont {C.}~\bibnamefont
  {Hwang}}, \bibinfo {author} {\bibfnamefont {Z.}~\bibnamefont {Hussain}},
  \bibinfo {author} {\bibfnamefont {Y.}~\bibnamefont {Chen}}, \bibinfo {author}
  {\bibfnamefont {M.~M.}\ \bibnamefont {Ugeda}}, \bibinfo {author}
  {\bibfnamefont {Z.}~\bibnamefont {Liu}}, \bibinfo {author} {\bibfnamefont
  {X.}~\bibnamefont {Xie}}, \bibinfo {author} {\bibfnamefont {T.~P.}\
  \bibnamefont {Devereaux}}, \bibinfo {author} {\bibfnamefont {M.~F.}\
  \bibnamefont {Crommie}}, \bibinfo {author} {\bibfnamefont {S.-K.}\
  \bibnamefont {Mo}}, \ and\ \bibinfo {author} {\bibfnamefont {Z.-X.}\
  \bibnamefont {Shen}},\ }\href {\doibase 10.1038/nphys4174} {\bibfield
  {journal} {\bibinfo  {journal} {Nature Physics}\ }\textbf {\bibinfo {volume}
  {13}},\ \bibinfo {pages} {683} (\bibinfo {year} {2017})}\BibitemShut
  {NoStop}%
\bibitem [{\citenamefont {Zareapour}\ \emph {et~al.}(2012)\citenamefont
  {Zareapour}, \citenamefont {Hayat}, \citenamefont {Zhao}, \citenamefont
  {Kreshchuk}, \citenamefont {Jain}, \citenamefont {Kwok}, \citenamefont {Lee},
  \citenamefont {Cheong}, \citenamefont {Xu}, \citenamefont {Yang},
  \citenamefont {Gu}, \citenamefont {Jia}, \citenamefont {Cava},\ and\
  \citenamefont {Burch}}]{Zareapour:2012ja}%
  \BibitemOpen
  \bibfield  {author} {\bibinfo {author} {\bibfnamefont {P.}~\bibnamefont
  {Zareapour}}, \bibinfo {author} {\bibfnamefont {A.}~\bibnamefont {Hayat}},
  \bibinfo {author} {\bibfnamefont {S.~Y.~F.}\ \bibnamefont {Zhao}}, \bibinfo
  {author} {\bibfnamefont {M.}~\bibnamefont {Kreshchuk}}, \bibinfo {author}
  {\bibfnamefont {A.}~\bibnamefont {Jain}}, \bibinfo {author} {\bibfnamefont
  {D.~C.}\ \bibnamefont {Kwok}}, \bibinfo {author} {\bibfnamefont
  {N.}~\bibnamefont {Lee}}, \bibinfo {author} {\bibfnamefont {S.-W.}\
  \bibnamefont {Cheong}}, \bibinfo {author} {\bibfnamefont {Z.}~\bibnamefont
  {Xu}}, \bibinfo {author} {\bibfnamefont {A.}~\bibnamefont {Yang}}, \bibinfo
  {author} {\bibfnamefont {G.~D.}\ \bibnamefont {Gu}}, \bibinfo {author}
  {\bibfnamefont {S.}~\bibnamefont {Jia}}, \bibinfo {author} {\bibfnamefont
  {R.~J.}\ \bibnamefont {Cava}}, \ and\ \bibinfo {author} {\bibfnamefont
  {K.~S.}\ \bibnamefont {Burch}},\ }\href@noop {} {\bibfield  {journal}
  {\bibinfo  {journal} {Nature communications}\ }\textbf {\bibinfo {volume}
  {3}},\ \bibinfo {pages} {1056} (\bibinfo {year} {2012})}\BibitemShut
  {NoStop}%
\bibitem [{\citenamefont {Island}\ \emph {et~al.}(2019)\citenamefont {Island},
  \citenamefont {Cui}, \citenamefont {Lewandowski}, \citenamefont {Khoo},
  \citenamefont {Spanton}, \citenamefont {Zhou}, \citenamefont {Rhodes},
  \citenamefont {Hone}, \citenamefont {Taniguchi}, \citenamefont {Watanabe},
  \citenamefont {Levitov}, \citenamefont {Zaletel},\ and\ \citenamefont
  {Young}}]{Island2019}%
  \BibitemOpen
  \bibfield  {author} {\bibinfo {author} {\bibfnamefont {J.~O.}\ \bibnamefont
  {Island}}, \bibinfo {author} {\bibfnamefont {X.}~\bibnamefont {Cui}},
  \bibinfo {author} {\bibfnamefont {C.}~\bibnamefont {Lewandowski}}, \bibinfo
  {author} {\bibfnamefont {J.~Y.}\ \bibnamefont {Khoo}}, \bibinfo {author}
  {\bibfnamefont {E.~M.}\ \bibnamefont {Spanton}}, \bibinfo {author}
  {\bibfnamefont {H.}~\bibnamefont {Zhou}}, \bibinfo {author} {\bibfnamefont
  {D.}~\bibnamefont {Rhodes}}, \bibinfo {author} {\bibfnamefont {J.~C.}\
  \bibnamefont {Hone}}, \bibinfo {author} {\bibfnamefont {T.}~\bibnamefont
  {Taniguchi}}, \bibinfo {author} {\bibfnamefont {K.}~\bibnamefont {Watanabe}},
  \bibinfo {author} {\bibfnamefont {L.~S.}\ \bibnamefont {Levitov}}, \bibinfo
  {author} {\bibfnamefont {M.~P.}\ \bibnamefont {Zaletel}}, \ and\ \bibinfo
  {author} {\bibfnamefont {A.~F.}\ \bibnamefont {Young}},\ }\href {\doibase
  10.1038/s41586-019-1304-2} {\bibfield  {journal}
  {\bibinfo  {journal} {Nature}\ }\textbf {\bibinfo {volume}
  {571}},\ \bibinfo {pages} {85}}  (\bibinfo {year} {2019})\BibitemShut {NoStop}%
\bibitem [{\citenamefont {Castellanos-Gomez}\ \emph {et~al.}(2014)\citenamefont
  {Castellanos-Gomez}, \citenamefont {Buscema}, \citenamefont {Molenaar},
  \citenamefont {Singh}, \citenamefont {Janssen}, \citenamefont {van~der
  Zant},\ and\ \citenamefont {Steele}}]{Castellanos_Gomez_2014}%
  \BibitemOpen
  \bibfield  {author} {\bibinfo {author} {\bibfnamefont {A.}~\bibnamefont
  {Castellanos-Gomez}}, \bibinfo {author} {\bibfnamefont {M.}~\bibnamefont
  {Buscema}}, \bibinfo {author} {\bibfnamefont {R.}~\bibnamefont {Molenaar}},
  \bibinfo {author} {\bibfnamefont {V.}~\bibnamefont {Singh}}, \bibinfo
  {author} {\bibfnamefont {L.}~\bibnamefont {Janssen}}, \bibinfo {author}
  {\bibfnamefont {H.~S.~J.}\ \bibnamefont {van~der Zant}}, \ and\ \bibinfo
  {author} {\bibfnamefont {G.~A.}\ \bibnamefont {Steele}},\ }\href {\doibase
  10.1088/2053-1583/1/1/011002} {\bibfield  {journal} {\bibinfo  {journal} {2D
  Materials}\ }\textbf {\bibinfo {volume} {1}},\ \bibinfo {pages} {011002}
  (\bibinfo {year} {2014})}\BibitemShut {NoStop}%
\bibitem [{\citenamefont {Brotons-Gisbert}\ \emph {et~al.}(2019)\citenamefont
  {Brotons-Gisbert}, \citenamefont {Branny}, \citenamefont {Kumar},
  \citenamefont {Picard}, \citenamefont {Proux}, \citenamefont {Gray},
  \citenamefont {Burch}, \citenamefont {Watanabe}, \citenamefont {Taniguchi},\
  and\ \citenamefont {Gerardot}}]{Brotons-Gisbert2019}%
  \BibitemOpen
  \bibfield  {author} {\bibinfo {author} {\bibfnamefont {M.}~\bibnamefont
  {Brotons-Gisbert}}, \bibinfo {author} {\bibfnamefont {A.}~\bibnamefont
  {Branny}}, \bibinfo {author} {\bibfnamefont {S.}~\bibnamefont {Kumar}},
  \bibinfo {author} {\bibfnamefont {R.}~\bibnamefont {Picard}}, \bibinfo
  {author} {\bibfnamefont {R.}~\bibnamefont {Proux}}, \bibinfo {author}
  {\bibfnamefont {M.}~\bibnamefont {Gray}}, \bibinfo {author} {\bibfnamefont
  {K.~S.}\ \bibnamefont {Burch}}, \bibinfo {author} {\bibfnamefont
  {K.}~\bibnamefont {Watanabe}}, \bibinfo {author} {\bibfnamefont
  {T.}~\bibnamefont {Taniguchi}}, \ and\ \bibinfo {author} {\bibfnamefont
  {B.~D.}\ \bibnamefont {Gerardot}},\ }\href {\doibase
  10.1038/s41565-019-0402-5} {\bibfield  {journal} {\bibinfo  {journal} {Nature
  Nanotechnology}\ }\textbf {\bibinfo {volume} {14}},\ \bibinfo {pages} {442}
  (\bibinfo {year} {2019})}\BibitemShut {NoStop}%
\bibitem [{\citenamefont {Gray}\ \emph {et~al.}(2019)\citenamefont {Gray},
  \citenamefont {Freudenstein}, \citenamefont {Zhao}, \citenamefont {O'Connor},
  \citenamefont {Jenkins}, \citenamefont {Kumar}, \citenamefont {Hoek},
  \citenamefont {Kopec}, \citenamefont {Huh}, \citenamefont {Taniguchi},
  \citenamefont {Watanabe}, \citenamefont {Zhong}, \citenamefont {Kim},
  \citenamefont {Gu},\ and\ \citenamefont {Burch}}]{Gray2019}%
  \BibitemOpen
  \bibfield  {author} {\bibinfo {author} {\bibfnamefont {M.~J.}\ \bibnamefont
  {Gray}}, \bibinfo {author} {\bibfnamefont {J.}~\bibnamefont {Freudenstein}},
  \bibinfo {author} {\bibfnamefont {S.~Y.~F.}\ \bibnamefont {Zhao}}, \bibinfo
  {author} {\bibfnamefont {R.}~\bibnamefont {O'Connor}}, \bibinfo {author}
  {\bibfnamefont {S.}~\bibnamefont {Jenkins}}, \bibinfo {author} {\bibfnamefont
  {N.}~\bibnamefont {Kumar}}, \bibinfo {author} {\bibfnamefont
  {M.}~\bibnamefont {Hoek}}, \bibinfo {author} {\bibfnamefont {A.}~\bibnamefont
  {Kopec}}, \bibinfo {author} {\bibfnamefont {S.}~\bibnamefont {Huh}}, \bibinfo
  {author} {\bibfnamefont {T.}~\bibnamefont {Taniguchi}}, \bibinfo {author}
  {\bibfnamefont {K.}~\bibnamefont {Watanabe}}, \bibinfo {author}
  {\bibfnamefont {R.}~\bibnamefont {Zhong}}, \bibinfo {author} {\bibfnamefont
  {C.}~\bibnamefont {Kim}}, \bibinfo {author} {\bibfnamefont {G.~D.}\
  \bibnamefont {Gu}}, \ and\ \bibinfo {author} {\bibfnamefont {K.~S.}\
  \bibnamefont {Burch}},\ }\href {\doibase 10.1021/acs.nanolett.9b00844}
  {\bibfield  {journal} {\bibinfo  {journal} {Nano Letters}\ }\textbf {\bibinfo
  {volume} {19}},\ \bibinfo {pages} {4890} (\bibinfo {year} {2019})}\BibitemShut {NoStop}%
\bibitem [{\citenamefont {Kumar}\ \emph {et~al.}(2020)\citenamefont {Kumar},
  \citenamefont {Wang}, \citenamefont {Ortiz-Marquez}, \citenamefont
  {Catalano}, \citenamefont {Gray}, \citenamefont {Biglari}, \citenamefont
  {Hikari}, \citenamefont {Ling}, \citenamefont {Gao}, \citenamefont {van
  Opijnen},\ and\ \citenamefont {Burch}}]{KUMAR2020112123}%
  \BibitemOpen
  \bibfield  {author} {\bibinfo {author} {\bibfnamefont {N.}~\bibnamefont
  {Kumar}}, \bibinfo {author} {\bibfnamefont {W.}~\bibnamefont {Wang}},
  \bibinfo {author} {\bibfnamefont {J.~C.}\ \bibnamefont {Ortiz-Marquez}},
  \bibinfo {author} {\bibfnamefont {M.}~\bibnamefont {Catalano}}, \bibinfo
  {author} {\bibfnamefont {M.}~\bibnamefont {Gray}}, \bibinfo {author}
  {\bibfnamefont {N.}~\bibnamefont {Biglari}}, \bibinfo {author} {\bibfnamefont
  {K.}~\bibnamefont {Hikari}}, \bibinfo {author} {\bibfnamefont
  {X.}~\bibnamefont {Ling}}, \bibinfo {author} {\bibfnamefont {J.}~\bibnamefont
  {Gao}}, \bibinfo {author} {\bibfnamefont {T.}~\bibnamefont {van Opijnen}}, \
  and\ \bibinfo {author} {\bibfnamefont {K.~S.}\ \bibnamefont {Burch}},\ }\href
  {\doibase https://doi.org/10.1016/j.bios.2020.112123} {\bibfield  {journal}
  {\bibinfo  {journal} {Biosensors and Bioelectronics}\ ,\ \textbf {\bibinfo
  {volume} {156}},\ \bibinfo {pages}
  {112123}} (\bibinfo {year} {2020})}\BibitemShut {NoStop}%
\bibitem [{\citenamefont {Zhao}\ \emph {et~al.}(2019)\citenamefont {Zhao},
  \citenamefont {Poccia}, \citenamefont {Panetta}, \citenamefont {Yu},
  \citenamefont {Johnson}, \citenamefont {Yoo}, \citenamefont {Zhong},
  \citenamefont {Gu}, \citenamefont {Watanabe}, \citenamefont {Taniguchi},
  \citenamefont {Postolova}, \citenamefont {Vinokur},\ and\ \citenamefont
  {Kim}}]{Zhao2019}%
  \BibitemOpen
  \bibfield  {author} {\bibinfo {author} {\bibfnamefont {S.~Y.~F.}\
  \bibnamefont {Zhao}}, \bibinfo {author} {\bibfnamefont {N.}~\bibnamefont
  {Poccia}}, \bibinfo {author} {\bibfnamefont {M.~G.}\ \bibnamefont {Panetta}},
  \bibinfo {author} {\bibfnamefont {C.}~\bibnamefont {Yu}}, \bibinfo {author}
  {\bibfnamefont {J.~W.}\ \bibnamefont {Johnson}}, \bibinfo {author}
  {\bibfnamefont {H.}~\bibnamefont {Yoo}}, \bibinfo {author} {\bibfnamefont
  {R.}~\bibnamefont {Zhong}}, \bibinfo {author} {\bibfnamefont {G.~D.}\
  \bibnamefont {Gu}}, \bibinfo {author} {\bibfnamefont {K.}~\bibnamefont
  {Watanabe}}, \bibinfo {author} {\bibfnamefont {T.}~\bibnamefont {Taniguchi}},
  \bibinfo {author} {\bibfnamefont {S.~V.}\ \bibnamefont {Postolova}}, \bibinfo
  {author} {\bibfnamefont {V.~M.}\ \bibnamefont {Vinokur}}, \ and\ \bibinfo
  {author} {\bibfnamefont {P.}~\bibnamefont {Kim}},\ }\href {\doibase
  10.1103/PhysRevLett.122.247001} {\bibfield  {journal} {\bibinfo  {journal}
  {Phys. Rev. Lett.}\ }\textbf {\bibinfo {volume} {122}},\ \bibinfo {pages}
  {247001} (\bibinfo {year} {2019})}\BibitemShut {NoStop}%
\bibitem [{\citenamefont {Zalic}\ \emph {et~al.}(2019)\citenamefont {Zalic},
  \citenamefont {Simon}, \citenamefont {Remennik}, \citenamefont {Vakahi},
  \citenamefont {Gu},\ and\ \citenamefont {Steinberg}}]{Zalic2019}%
  \BibitemOpen
  \bibfield  {author} {\bibinfo {author} {\bibfnamefont {A.}~\bibnamefont
  {Zalic}}, \bibinfo {author} {\bibfnamefont {S.}~\bibnamefont {Simon}},
  \bibinfo {author} {\bibfnamefont {S.}~\bibnamefont {Remennik}}, \bibinfo
  {author} {\bibfnamefont {A.}~\bibnamefont {Vakahi}}, \bibinfo {author}
  {\bibfnamefont {G.~D.}\ \bibnamefont {Gu}}, \ and\ \bibinfo {author}
  {\bibfnamefont {H.}~\bibnamefont {Steinberg}},\ }\href {\doibase
  10.1103/PhysRevB.100.064517} {\bibfield  {journal} {\bibinfo  {journal}
  {Phys. Rev. B}\ }\textbf {\bibinfo {volume} {100}},\ \bibinfo {pages}
  {064517} (\bibinfo {year} {2019})}\BibitemShut {NoStop}%
\bibitem [{\citenamefont {Damasco}\ \emph {et~al.}(2019)\citenamefont
  {Damasco}, \citenamefont {Gill}, \citenamefont {Gazibegovic}, \citenamefont
  {Badawy}, \citenamefont {Bakkers},\ and\ \citenamefont
  {Mason}}]{Damasco2019tunnel}%
  \BibitemOpen
  \bibfield  {author} {\bibinfo {author} {\bibfnamefont {J.}~\bibnamefont
  {Damasco}}, \bibinfo {author} {\bibfnamefont {S.~T.}\ \bibnamefont {Gill}},
  \bibinfo {author} {\bibfnamefont {S.}~\bibnamefont {Gazibegovic}}, \bibinfo
  {author} {\bibfnamefont {G.}~\bibnamefont {Badawy}}, \bibinfo {author}
  {\bibfnamefont {E.~P. A.~M.}\ \bibnamefont {Bakkers}}, \ and\ \bibinfo
  {author} {\bibfnamefont {N.}~\bibnamefont {Mason}},\ }\href {\doibase
  10.1063/1.5108539} {\bibfield  {journal} {\bibinfo  {journal} {Applied
  Physics Letters}\ }\textbf {\bibinfo {volume} {115}},\ \bibinfo {pages}
  {043503} (\bibinfo {year} {2019})}\BibitemShut {NoStop}%
\bibitem [{\citenamefont {Ping}\ \emph {et~al.}(2016)\citenamefont {Ping},
  \citenamefont {Vishnubhotla}, \citenamefont {Vrudhula},\ and\ \citenamefont
  {Johnson}}]{doi:10.1021/acsnano.6b04110}%
  \BibitemOpen
  \bibfield  {author} {\bibinfo {author} {\bibfnamefont {J.}~\bibnamefont
  {Ping}}, \bibinfo {author} {\bibfnamefont {R.}~\bibnamefont {Vishnubhotla}},
  \bibinfo {author} {\bibfnamefont {A.}~\bibnamefont {Vrudhula}}, \ and\
  \bibinfo {author} {\bibfnamefont {A.~T.~C.}\ \bibnamefont {Johnson}},\ }\href
  {\doibase 10.1021/acsnano.6b04110} {\bibfield  {journal} {\bibinfo  {journal}
  {ACS Nano}\ }\textbf {\bibinfo {volume} {10}},\ \bibinfo {pages} {8700}
  (\bibinfo {year} {2016})}\BibitemShut {NoStop}%
\bibitem [{\citenamefont {Gerber}\ \emph {et~al.}(2017)\citenamefont {Gerber},
  \citenamefont {Yang}, \citenamefont {Zhu}, \citenamefont {Soifer},
  \citenamefont {Sobota}, \citenamefont {Rebec}, \citenamefont {Lee},
  \citenamefont {Jia}, \citenamefont {Moritz}, \citenamefont {Jia},
  \citenamefont {Gauthier}, \citenamefont {Li}, \citenamefont {Leuenberger},
  \citenamefont {Zhang}, \citenamefont {Chaix}, \citenamefont {Li},
  \citenamefont {Jang}, \citenamefont {Lee}, \citenamefont {Yi}, \citenamefont
  {Dakovski}, \citenamefont {Song}, \citenamefont {Glownia}, \citenamefont
  {Nelson}, \citenamefont {Kim}, \citenamefont {Chuang}, \citenamefont
  {Hussain}, \citenamefont {Moore}, \citenamefont {Devereaux}, \citenamefont
  {Lee}, \citenamefont {Kirchmann},\ and\ \citenamefont
  {Shen}}]{Gerber2017FeSe}%
  \BibitemOpen
  \bibfield  {author} {\bibinfo {author} {\bibfnamefont {S.}~\bibnamefont
  {Gerber}}, \bibinfo {author} {\bibfnamefont {S.-L.}\ \bibnamefont {Yang}},
  \bibinfo {author} {\bibfnamefont {D.}~\bibnamefont {Zhu}}, \bibinfo {author}
  {\bibfnamefont {H.}~\bibnamefont {Soifer}}, \bibinfo {author} {\bibfnamefont
  {J.~A.}\ \bibnamefont {Sobota}}, \bibinfo {author} {\bibfnamefont
  {S.}~\bibnamefont {Rebec}}, \bibinfo {author} {\bibfnamefont {J.~J.}\
  \bibnamefont {Lee}}, \bibinfo {author} {\bibfnamefont {T.}~\bibnamefont
  {Jia}}, \bibinfo {author} {\bibfnamefont {B.}~\bibnamefont {Moritz}},
  \bibinfo {author} {\bibfnamefont {C.}~\bibnamefont {Jia}}, \bibinfo {author}
  {\bibfnamefont {A.}~\bibnamefont {Gauthier}}, \bibinfo {author}
  {\bibfnamefont {Y.}~\bibnamefont {Li}}, \bibinfo {author} {\bibfnamefont
  {D.}~\bibnamefont {Leuenberger}}, \bibinfo {author} {\bibfnamefont
  {Y.}~\bibnamefont {Zhang}}, \bibinfo {author} {\bibfnamefont
  {L.}~\bibnamefont {Chaix}}, \bibinfo {author} {\bibfnamefont
  {W.}~\bibnamefont {Li}}, \bibinfo {author} {\bibfnamefont {H.}~\bibnamefont
  {Jang}}, \bibinfo {author} {\bibfnamefont {J.-S.}\ \bibnamefont {Lee}},
  \bibinfo {author} {\bibfnamefont {M.}~\bibnamefont {Yi}}, \bibinfo {author}
  {\bibfnamefont {G.~L.}\ \bibnamefont {Dakovski}}, \bibinfo {author}
  {\bibfnamefont {S.}~\bibnamefont {Song}}, \bibinfo {author} {\bibfnamefont
  {J.~M.}\ \bibnamefont {Glownia}}, \bibinfo {author} {\bibfnamefont
  {S.}~\bibnamefont {Nelson}}, \bibinfo {author} {\bibfnamefont {K.~W.}\
  \bibnamefont {Kim}}, \bibinfo {author} {\bibfnamefont {Y.-D.}\ \bibnamefont
  {Chuang}}, \bibinfo {author} {\bibfnamefont {Z.}~\bibnamefont {Hussain}},
  \bibinfo {author} {\bibfnamefont {R.~G.}\ \bibnamefont {Moore}}, \bibinfo
  {author} {\bibfnamefont {T.~P.}\ \bibnamefont {Devereaux}}, \bibinfo {author}
  {\bibfnamefont {W.-S.}\ \bibnamefont {Lee}}, \bibinfo {author} {\bibfnamefont
  {P.~S.}\ \bibnamefont {Kirchmann}}, \ and\ \bibinfo {author} {\bibfnamefont
  {Z.-X.}\ \bibnamefont {Shen}},\ }\href {\doibase 10.1126/science.aak9946}
  {\bibfield  {journal} {\bibinfo  {journal} {Science}\ }\textbf {\bibinfo
  {volume} {357}},\ \bibinfo {pages} {71} (\bibinfo {year} {2017})}\BibitemShut
  {NoStop}%
\bibitem [{\citenamefont {Wang}, \citenamefont {Ma},\ and\ \citenamefont
  {Xue}(2016)}]{2016SuScT..29l3001W}%
  \BibitemOpen
  \bibfield  {author} {\bibinfo {author} {\bibfnamefont {L.}~\bibnamefont
  {Wang}}, \bibinfo {author} {\bibfnamefont {X.}~\bibnamefont {Ma}}, \ and\
  \bibinfo {author} {\bibfnamefont {Q.-K.}\ \bibnamefont {Xue}},\ }\href@noop
  {} {\bibfield  {journal} {\bibinfo  {journal} {Superconductor Science and
  Technology}\ }\textbf {\bibinfo {volume} {29}},\ \bibinfo {pages} {123001}
  (\bibinfo {year} {2016})}\BibitemShut {NoStop}%
\bibitem [{\citenamefont {Hellman}\ \emph {et~al.}(2017)\citenamefont
  {Hellman}, \citenamefont {Hoffmann}, \citenamefont {Tserkovnyak},
  \citenamefont {Beach}, \citenamefont {Fullerton}, \citenamefont {Leighton},
  \citenamefont {MacDonald}, \citenamefont {Ralph}, \citenamefont {Arena},
  \citenamefont {D\"urr}, \citenamefont {Fischer}, \citenamefont {Grollier},
  \citenamefont {Heremans}, \citenamefont {Jungwirth}, \citenamefont {Kimel},
  \citenamefont {Koopmans}, \citenamefont {Krivorotov}, \citenamefont {May},
  \citenamefont {Petford-Long}, \citenamefont {Rondinelli}, \citenamefont
  {Samarth}, \citenamefont {Schuller}, \citenamefont {Slavin}, \citenamefont
  {Stiles}, \citenamefont {Tchernyshyov}, \citenamefont {Thiaville},\ and\
  \citenamefont {Zink}}]{Hellman2017Rev}%
  \BibitemOpen
  \bibfield  {author} {\bibinfo {author} {\bibfnamefont {F.}~\bibnamefont
  {Hellman}}, \bibinfo {author} {\bibfnamefont {A.}~\bibnamefont {Hoffmann}},
  \bibinfo {author} {\bibfnamefont {Y.}~\bibnamefont {Tserkovnyak}}, \bibinfo
  {author} {\bibfnamefont {G.~S.~D.}\ \bibnamefont {Beach}}, \bibinfo {author}
  {\bibfnamefont {E.~E.}\ \bibnamefont {Fullerton}}, \bibinfo {author}
  {\bibfnamefont {C.}~\bibnamefont {Leighton}}, \bibinfo {author}
  {\bibfnamefont {A.~H.}\ \bibnamefont {MacDonald}}, \bibinfo {author}
  {\bibfnamefont {D.~C.}\ \bibnamefont {Ralph}}, \bibinfo {author}
  {\bibfnamefont {D.~A.}\ \bibnamefont {Arena}}, \bibinfo {author}
  {\bibfnamefont {H.~A.}\ \bibnamefont {D\"urr}}, \bibinfo {author}
  {\bibfnamefont {P.}~\bibnamefont {Fischer}}, \bibinfo {author} {\bibfnamefont
  {J.}~\bibnamefont {Grollier}}, \bibinfo {author} {\bibfnamefont {J.~P.}\
  \bibnamefont {Heremans}}, \bibinfo {author} {\bibfnamefont {T.}~\bibnamefont
  {Jungwirth}}, \bibinfo {author} {\bibfnamefont {A.~V.}\ \bibnamefont
  {Kimel}}, \bibinfo {author} {\bibfnamefont {B.}~\bibnamefont {Koopmans}},
  \bibinfo {author} {\bibfnamefont {I.~N.}\ \bibnamefont {Krivorotov}},
  \bibinfo {author} {\bibfnamefont {S.~J.}\ \bibnamefont {May}}, \bibinfo
  {author} {\bibfnamefont {A.~K.}\ \bibnamefont {Petford-Long}}, \bibinfo
  {author} {\bibfnamefont {J.~M.}\ \bibnamefont {Rondinelli}}, \bibinfo
  {author} {\bibfnamefont {N.}~\bibnamefont {Samarth}}, \bibinfo {author}
  {\bibfnamefont {I.~K.}\ \bibnamefont {Schuller}}, \bibinfo {author}
  {\bibfnamefont {A.~N.}\ \bibnamefont {Slavin}}, \bibinfo {author}
  {\bibfnamefont {M.~D.}\ \bibnamefont {Stiles}}, \bibinfo {author}
  {\bibfnamefont {O.}~\bibnamefont {Tchernyshyov}}, \bibinfo {author}
  {\bibfnamefont {A.}~\bibnamefont {Thiaville}}, \ and\ \bibinfo {author}
  {\bibfnamefont {B.~L.}\ \bibnamefont {Zink}},\ }\href {\doibase
  10.1103/RevModPhys.89.025006} {\bibfield  {journal} {\bibinfo  {journal}
  {Rev. Mod. Phys.}\ }\textbf {\bibinfo {volume} {89}},\ \bibinfo {pages}
  {025006} (\bibinfo {year} {2017})}\BibitemShut {NoStop}%
\bibitem [{\citenamefont {Woods}\ \emph {et~al.}(2014)\citenamefont {Woods},
  \citenamefont {Britnell}, \citenamefont {Eckmann}, \citenamefont {Yu},
  \citenamefont {Gorbachev}, \citenamefont {Kretinin}, \citenamefont {Park},
  \citenamefont {Ponomarenko}, \citenamefont {Katsnelson}, \citenamefont
  {Gornostyrev}, \citenamefont {Watanabe}, \citenamefont {Taniguchi},
  \citenamefont {Casiraghi}, \citenamefont {Guo}, \citenamefont {Geim},\ and\
  \citenamefont {Novoselov}}]{Woods2014}%
  \BibitemOpen
  \bibfield  {author} {\bibinfo {author} {\bibfnamefont {C.}~\bibnamefont
  {Woods}}, \bibinfo {author} {\bibfnamefont {L.}~\bibnamefont {Britnell}},
  \bibinfo {author} {\bibfnamefont {A.}~\bibnamefont {Eckmann}}, \bibinfo
  {author} {\bibfnamefont {G.}~\bibnamefont {Yu}}, \bibinfo {author}
  {\bibfnamefont {R.}~\bibnamefont {Gorbachev}}, \bibinfo {author}
  {\bibfnamefont {A.}~\bibnamefont {Kretinin}}, \bibinfo {author}
  {\bibfnamefont {J.}~\bibnamefont {Park}}, \bibinfo {author} {\bibfnamefont
  {L.}~\bibnamefont {Ponomarenko}}, \bibinfo {author} {\bibfnamefont
  {M.}~\bibnamefont {Katsnelson}}, \bibinfo {author} {\bibfnamefont
  {Y.}~\bibnamefont {Gornostyrev}}, \bibinfo {author} {\bibfnamefont
  {K.}~\bibnamefont {Watanabe}}, \bibinfo {author} {\bibfnamefont
  {T.}~\bibnamefont {Taniguchi}}, \bibinfo {author} {\bibfnamefont
  {C.}~\bibnamefont {Casiraghi}}, \bibinfo {author} {\bibfnamefont
  {H.}~\bibnamefont {Guo}}, \bibinfo {author} {\bibfnamefont {A.}~\bibnamefont
  {Geim}}, \ and\ \bibinfo {author} {\bibfnamefont {K.}~\bibnamefont
  {Novoselov}},\ }\href {\doibase 10.1038/nphys2954} {\bibfield  {journal}
  {\bibinfo  {journal} {Nature Physics}\ }\textbf {\bibinfo {volume} {10}},\
  \bibinfo {pages} {451} (\bibinfo {year} {2014})}\BibitemShut {NoStop}%
\bibitem [{\citenamefont {Tran}\ \emph {et~al.}(2019)\citenamefont {Tran},
  \citenamefont {Moody}, \citenamefont {Wu}, \citenamefont {Lu}, \citenamefont
  {Choi}, \citenamefont {Kim}, \citenamefont {Rai}, \citenamefont {Sanchez},
  \citenamefont {Quan}, \citenamefont {Singh}, \citenamefont {Embley},
  \citenamefont {Zepeda}, \citenamefont {Campbell}, \citenamefont {Autry},
  \citenamefont {Taniguchi}, \citenamefont {Watanabe}, \citenamefont {Lu},
  \citenamefont {Banerjee}, \citenamefont {Silverman}, \citenamefont {Kim},
  \citenamefont {Tutuc}, \citenamefont {Yang}, \citenamefont {MacDonald},\ and\
  \citenamefont {Li}}]{Tran:2019aa}%
  \BibitemOpen
  \bibfield  {author} {\bibinfo {author} {\bibfnamefont {K.}~\bibnamefont
  {Tran}}, \bibinfo {author} {\bibfnamefont {G.}~\bibnamefont {Moody}},
  \bibinfo {author} {\bibfnamefont {F.}~\bibnamefont {Wu}}, \bibinfo {author}
  {\bibfnamefont {X.}~\bibnamefont {Lu}}, \bibinfo {author} {\bibfnamefont
  {J.}~\bibnamefont {Choi}}, \bibinfo {author} {\bibfnamefont {K.}~\bibnamefont
  {Kim}}, \bibinfo {author} {\bibfnamefont {A.}~\bibnamefont {Rai}}, \bibinfo
  {author} {\bibfnamefont {D.~A.}\ \bibnamefont {Sanchez}}, \bibinfo {author}
  {\bibfnamefont {J.}~\bibnamefont {Quan}}, \bibinfo {author} {\bibfnamefont
  {A.}~\bibnamefont {Singh}}, \bibinfo {author} {\bibfnamefont
  {J.}~\bibnamefont {Embley}}, \bibinfo {author} {\bibfnamefont
  {A.}~\bibnamefont {Zepeda}}, \bibinfo {author} {\bibfnamefont
  {M.}~\bibnamefont {Campbell}}, \bibinfo {author} {\bibfnamefont
  {T.}~\bibnamefont {Autry}}, \bibinfo {author} {\bibfnamefont
  {T.}~\bibnamefont {Taniguchi}}, \bibinfo {author} {\bibfnamefont
  {K.}~\bibnamefont {Watanabe}}, \bibinfo {author} {\bibfnamefont
  {N.}~\bibnamefont {Lu}}, \bibinfo {author} {\bibfnamefont {S.~K.}\
  \bibnamefont {Banerjee}}, \bibinfo {author} {\bibfnamefont {K.~L.}\
  \bibnamefont {Silverman}}, \bibinfo {author} {\bibfnamefont {S.}~\bibnamefont
  {Kim}}, \bibinfo {author} {\bibfnamefont {E.}~\bibnamefont {Tutuc}}, \bibinfo
  {author} {\bibfnamefont {L.}~\bibnamefont {Yang}}, \bibinfo {author}
  {\bibfnamefont {A.~H.}\ \bibnamefont {MacDonald}}, \ and\ \bibinfo {author}
  {\bibfnamefont {X.}~\bibnamefont {Li}},\ }\href {\doibase
  10.1038/s41586-019-0975-z} {\bibfield  {journal} {\bibinfo  {journal}
  {Nature}\ }\textbf {\bibinfo {volume} {567}},\ \bibinfo {pages} {71}
  (\bibinfo {year} {2019})}\BibitemShut {NoStop}%
\bibitem [{\citenamefont {Jin}\ \emph {et~al.}(2019)\citenamefont {Jin},
  \citenamefont {Regan}, \citenamefont {Yan}, \citenamefont {Iqbal
  Bakti~Utama}, \citenamefont {Wang}, \citenamefont {Zhao}, \citenamefont
  {Qin}, \citenamefont {Yang}, \citenamefont {Zheng}, \citenamefont {Shi},
  \citenamefont {Watanabe}, \citenamefont {Taniguchi}, \citenamefont {Tongay},
  \citenamefont {Zettl},\ and\ \citenamefont {Wang}}]{Jin:2019aa}%
  \BibitemOpen
  \bibfield  {author} {\bibinfo {author} {\bibfnamefont {C.}~\bibnamefont
  {Jin}}, \bibinfo {author} {\bibfnamefont {E.~C.}\ \bibnamefont {Regan}},
  \bibinfo {author} {\bibfnamefont {A.}~\bibnamefont {Yan}}, \bibinfo {author}
  {\bibfnamefont {M.}~\bibnamefont {Iqbal Bakti~Utama}}, \bibinfo {author}
  {\bibfnamefont {D.}~\bibnamefont {Wang}}, \bibinfo {author} {\bibfnamefont
  {S.}~\bibnamefont {Zhao}}, \bibinfo {author} {\bibfnamefont {Y.}~\bibnamefont
  {Qin}}, \bibinfo {author} {\bibfnamefont {S.}~\bibnamefont {Yang}}, \bibinfo
  {author} {\bibfnamefont {Z.}~\bibnamefont {Zheng}}, \bibinfo {author}
  {\bibfnamefont {S.}~\bibnamefont {Shi}}, \bibinfo {author} {\bibfnamefont
  {K.}~\bibnamefont {Watanabe}}, \bibinfo {author} {\bibfnamefont
  {T.}~\bibnamefont {Taniguchi}}, \bibinfo {author} {\bibfnamefont
  {S.}~\bibnamefont {Tongay}}, \bibinfo {author} {\bibfnamefont
  {A.}~\bibnamefont {Zettl}}, \ and\ \bibinfo {author} {\bibfnamefont
  {F.}~\bibnamefont {Wang}},\ }\href {\doibase 10.1038/s41586-019-0976-y}
  {\bibfield  {journal} {\bibinfo  {journal} {Nature}\ }\textbf {\bibinfo
  {volume} {567}},\ \bibinfo {pages} {76} (\bibinfo {year} {2019})}\BibitemShut
  {NoStop}%
\bibitem [{\citenamefont {Alexeev}\ \emph {et~al.}(2019)\citenamefont
  {Alexeev}, \citenamefont {Ruiz-Tijerina}, \citenamefont {Danovich},
  \citenamefont {Hamer}, \citenamefont {Terry}, \citenamefont {Nayak},
  \citenamefont {Ahn}, \citenamefont {Pak}, \citenamefont {Lee}, \citenamefont
  {Sohn}, \citenamefont {Molas}, \citenamefont {Koperski}, \citenamefont
  {Watanabe}, \citenamefont {Taniguchi}, \citenamefont {Novoselov},
  \citenamefont {Gorbachev}, \citenamefont {Shin}, \citenamefont {Fal'ko},\
  and\ \citenamefont {Tartakovskii}}]{Alexeev:2019aa}%
  \BibitemOpen
  \bibfield  {author} {\bibinfo {author} {\bibfnamefont {E.~M.}\ \bibnamefont
  {Alexeev}}, \bibinfo {author} {\bibfnamefont {D.~A.}\ \bibnamefont
  {Ruiz-Tijerina}}, \bibinfo {author} {\bibfnamefont {M.}~\bibnamefont
  {Danovich}}, \bibinfo {author} {\bibfnamefont {M.~J.}\ \bibnamefont {Hamer}},
  \bibinfo {author} {\bibfnamefont {D.~J.}\ \bibnamefont {Terry}}, \bibinfo
  {author} {\bibfnamefont {P.~K.}\ \bibnamefont {Nayak}}, \bibinfo {author}
  {\bibfnamefont {S.}~\bibnamefont {Ahn}}, \bibinfo {author} {\bibfnamefont
  {S.}~\bibnamefont {Pak}}, \bibinfo {author} {\bibfnamefont {J.}~\bibnamefont
  {Lee}}, \bibinfo {author} {\bibfnamefont {J.~I.}\ \bibnamefont {Sohn}},
  \bibinfo {author} {\bibfnamefont {M.~R.}\ \bibnamefont {Molas}}, \bibinfo
  {author} {\bibfnamefont {M.}~\bibnamefont {Koperski}}, \bibinfo {author}
  {\bibfnamefont {K.}~\bibnamefont {Watanabe}}, \bibinfo {author}
  {\bibfnamefont {T.}~\bibnamefont {Taniguchi}}, \bibinfo {author}
  {\bibfnamefont {K.~S.}\ \bibnamefont {Novoselov}}, \bibinfo {author}
  {\bibfnamefont {R.~V.}\ \bibnamefont {Gorbachev}}, \bibinfo {author}
  {\bibfnamefont {H.~S.}\ \bibnamefont {Shin}}, \bibinfo {author}
  {\bibfnamefont {V.~I.}\ \bibnamefont {Fal'ko}}, \ and\ \bibinfo {author}
  {\bibfnamefont {A.~I.}\ \bibnamefont {Tartakovskii}},\ }\href {\doibase
  10.1038/s41586-019-0986-9} {\bibfield  {journal} {\bibinfo  {journal}
  {Nature}\ }\textbf {\bibinfo {volume} {567}},\ \bibinfo {pages} {81}
  (\bibinfo {year} {2019})}\BibitemShut {NoStop}%
\bibitem [{\citenamefont {Yankowitz}\ \emph {et~al.}(2019)\citenamefont
  {Yankowitz}, \citenamefont {Chen}, \citenamefont {Polshyn}, \citenamefont
  {Zhang}, \citenamefont {Watanabe}, \citenamefont {Taniguchi}, \citenamefont
  {Graf}, \citenamefont {Young},\ and\ \citenamefont {Dean}}]{Yankowitz2019}%
  \BibitemOpen
  \bibfield  {author} {\bibinfo {author} {\bibfnamefont {M.}~\bibnamefont
  {Yankowitz}}, \bibinfo {author} {\bibfnamefont {S.}~\bibnamefont {Chen}},
  \bibinfo {author} {\bibfnamefont {H.}~\bibnamefont {Polshyn}}, \bibinfo
  {author} {\bibfnamefont {Y.}~\bibnamefont {Zhang}}, \bibinfo {author}
  {\bibfnamefont {K.}~\bibnamefont {Watanabe}}, \bibinfo {author}
  {\bibfnamefont {T.}~\bibnamefont {Taniguchi}}, \bibinfo {author}
  {\bibfnamefont {D.}~\bibnamefont {Graf}}, \bibinfo {author} {\bibfnamefont
  {A.~F.}\ \bibnamefont {Young}}, \ and\ \bibinfo {author} {\bibfnamefont
  {C.~R.}\ \bibnamefont {Dean}},\ }\href {\doibase 10.1126/science.aav1910}
  {\bibfield  {journal} {\bibinfo  {journal}
  {Science}\ }\textbf {\bibinfo {volume} {363}},\ \bibinfo {pages} {1059}
  (\bibinfo {year} {2019})}\BibitemShut {NoStop}
  \BibitemShut {NoStop}%
\bibitem [{\citenamefont {Cao}\ \emph {et~al.}(2018)\citenamefont {Cao},
  \citenamefont {Fatemi}, \citenamefont {Fang}, \citenamefont {Watanabe},
  \citenamefont {Taniguchi}, \citenamefont {Kaxiras},\ and\ \citenamefont
  {Jarillo-Herrero}}]{Cao2018}%
  \BibitemOpen
  \bibfield  {author} {\bibinfo {author} {\bibfnamefont {Y.}~\bibnamefont
  {Cao}}, \bibinfo {author} {\bibfnamefont {V.}~\bibnamefont {Fatemi}},
  \bibinfo {author} {\bibfnamefont {S.}~\bibnamefont {Fang}}, \bibinfo {author}
  {\bibfnamefont {K.}~\bibnamefont {Watanabe}}, \bibinfo {author}
  {\bibfnamefont {T.}~\bibnamefont {Taniguchi}}, \bibinfo {author}
  {\bibfnamefont {E.}~\bibnamefont {Kaxiras}}, \ and\ \bibinfo {author}
  {\bibfnamefont {P.}~\bibnamefont {Jarillo-Herrero}},\ }\href {\doibase
  10.1038/nature26160} {\bibfield  {journal} {\bibinfo  {journal}
  {Nature}\ }\textbf {\bibinfo {volume} {556}},\ \bibinfo {pages} {43}
  (\bibinfo {year} {2018})}\BibitemShut {NoStop}\BibitemShut {NoStop}%
\bibitem [{\citenamefont {{Yang}}\ \emph {et~al.}(2019)\citenamefont {{Yang}},
  \citenamefont {{Luo}}, \citenamefont {{Chi}}, \citenamefont {{Bonn}},\ and\
  \citenamefont {{Xia}}}]{Yang2019}%
  \BibitemOpen
  \bibfield  {author} {\bibinfo {author} {\bibfnamefont {R.}~\bibnamefont
  {{Yang}}}, \bibinfo {author} {\bibfnamefont {W.}~\bibnamefont {{Luo}}},
  \bibinfo {author} {\bibfnamefont {S.}~\bibnamefont {{Chi}}}, \bibinfo
  {author} {\bibfnamefont {D.}~\bibnamefont {{Bonn}}}, \ and\ \bibinfo {author}
  {\bibfnamefont {G.~M.}\ \bibnamefont {{Xia}}},\ }\href {\doibase
  10.1109/TNANO.2018.2877524} {\bibfield  {journal} {\bibinfo  {journal} {IEEE
  Transactions on Nanotechnology}\ }\textbf {\bibinfo {volume} {18}},\ \bibinfo
  {pages} {37} (\bibinfo {year} {2019})}\BibitemShut {NoStop}%
\bibitem [{\citenamefont {Kawai}, \citenamefont {Nabeshima},\ and\
  \citenamefont {Maeda}(2018)}]{Kawai_2018}%
  \BibitemOpen
  \bibfield  {author} {\bibinfo {author} {\bibfnamefont {M.}~\bibnamefont
  {Kawai}}, \bibinfo {author} {\bibfnamefont {F.}~\bibnamefont {Nabeshima}}, \
  and\ \bibinfo {author} {\bibfnamefont {A.}~\bibnamefont {Maeda}},\ }\href
  {\doibase 10.1088/1742-6596/1054/1/012023} {\bibfield  {journal} {\bibinfo
  {journal} {Journal of Physics: Conference Series}\ }\textbf {\bibinfo
  {volume} {1054}},\ \bibinfo {pages} {012023} (\bibinfo {year}
  {2018})}\BibitemShut {NoStop}%
\bibitem [{\citenamefont {Tian}\ \emph
  {et~al.}(2016{\natexlab{b}})\citenamefont {Tian}, \citenamefont {Reijnders},
  \citenamefont {Osterhoudt}, \citenamefont {Valmianski}, \citenamefont
  {Ramirez}, \citenamefont {Urban}, \citenamefont {Zhong}, \citenamefont
  {Schneeloch}, \citenamefont {Gu}, \citenamefont {Henslee},\ and\
  \citenamefont {Burch}}]{Tian2016}%
  \BibitemOpen
  \bibfield  {author} {\bibinfo {author} {\bibfnamefont {Y.}~\bibnamefont
  {Tian}}, \bibinfo {author} {\bibfnamefont {A.~A.}\ \bibnamefont {Reijnders}},
  \bibinfo {author} {\bibfnamefont {G.~B.}\ \bibnamefont {Osterhoudt}},
  \bibinfo {author} {\bibfnamefont {I.}~\bibnamefont {Valmianski}}, \bibinfo
  {author} {\bibfnamefont {J.~G.}\ \bibnamefont {Ramirez}}, \bibinfo {author}
  {\bibfnamefont {C.}~\bibnamefont {Urban}}, \bibinfo {author} {\bibfnamefont
  {R.}~\bibnamefont {Zhong}}, \bibinfo {author} {\bibfnamefont
  {J.}~\bibnamefont {Schneeloch}}, \bibinfo {author} {\bibfnamefont
  {G.}~\bibnamefont {Gu}}, \bibinfo {author} {\bibfnamefont {I.}~\bibnamefont
  {Henslee}}, \ and\ \bibinfo {author} {\bibfnamefont {K.~S.}\ \bibnamefont
  {Burch}},\ }\href {\doibase 10.1063/1.4944559} {\bibfield  {journal}
  {\bibinfo  {journal} {Review of Scientific Instruments}\ }\textbf {\bibinfo
  {volume} {87}},\ \bibinfo {pages} {043105} (\bibinfo {year}
  {2016}{\natexlab{b}})}\BibitemShut {NoStop}%
\bibitem [{\citenamefont {Firpo}\ \emph {et~al.}(2005)\citenamefont {Firpo},
  \citenamefont {Buatier~de Mongeot}, \citenamefont {Boragno},\ and\
  \citenamefont {Valbusa}}]{Firpo2005}%
  \BibitemOpen
  \bibfield  {author} {\bibinfo {author} {\bibfnamefont {G.}~\bibnamefont
  {Firpo}}, \bibinfo {author} {\bibfnamefont {F.}~\bibnamefont {Buatier~de
  Mongeot}}, \bibinfo {author} {\bibfnamefont {C.}~\bibnamefont {Boragno}}, \
  and\ \bibinfo {author} {\bibfnamefont {U.}~\bibnamefont {Valbusa}},\ }\href
  {\doibase 10.1063/1.1834493} {\bibfield  {journal} {\bibinfo  {journal}
  {Review of Scientific Instruments}\ }\textbf {\bibinfo {volume} {76}},\
  \bibinfo {pages} {026108} (\bibinfo {year} {2005})}\BibitemShut {NoStop}%
\bibitem [{\citenamefont {Tian}\ \emph
  {et~al.}(2016{\natexlab{c}})\citenamefont {Tian}, \citenamefont {Gray},
  \citenamefont {Ji}, \citenamefont {Cava},\ and\ \citenamefont
  {Burch}}]{YaoCGT2Dmat}%
  \BibitemOpen
  \bibfield  {author} {\bibinfo {author} {\bibfnamefont {Y.}~\bibnamefont
  {Tian}}, \bibinfo {author} {\bibfnamefont {M.~J.}\ \bibnamefont {Gray}},
  \bibinfo {author} {\bibfnamefont {H.}~\bibnamefont {Ji}}, \bibinfo {author}
  {\bibfnamefont {R.~J.}\ \bibnamefont {Cava}}, \ and\ \bibinfo {author}
  {\bibfnamefont {K.~S.}\ \bibnamefont {Burch}},\ }\href {https://is.gd/uprJmM}
  {\bibfield  {journal} {\bibinfo  {journal} {2D Materials}\ }\textbf {\bibinfo
  {volume} {3}},\ \bibinfo {pages} {025035} (\bibinfo {year}
  {2016}{\natexlab{c}})}\BibitemShut {NoStop}%
\end{thebibliography}


\end{document}